\documentclass[twocolumn]{autart}

\usepackage{mathtools}
\usepackage{amsmath}
\usepackage{amssymb}
\usepackage{mathrsfs}
\usepackage{graphicx}
\usepackage{subfigure}
\usepackage{bm}
\usepackage{hyperref}
\usepackage{enumerate}
\usepackage{color}
\usepackage{array}

\allowdisplaybreaks[4]

\graphicspath{{figures/}}

\newtheorem{asmp}{Assumption}
\newtheorem{lemm}{Lemma}

\newtheorem{rmk}{Remark}
\newtheorem{corr}{Corollary}

\newcommand{\mtcc}{\mathcal{C}}

\newcommand{\mtce}{\mathcal{E}}

\newcommand{\mtcg}{\mathcal{G}}

\newcommand{\mtcs}{\mathcal{S}}

\newcommand{\mtcv}{\mathcal{V}}

\newcommand{\bfl}{\mathbf{1}}
\newcommand{\bfo}{\mathbf{0}}
\newcommand{\bfz}{z}
\newcommand{\bfe}{e}

\newcommand{\RR}{\mathbb{R}}
\newcommand{\EE}{\mathbb{E}}
\newcommand{\PP}{\mathbb{P}}
\newcommand{\NN}{\mathbb{N}}
\newcommand{\eps}{\varepsilon}

\newcommand{\Let}{: =}
\newcommand{\teL}{= :}

\begin{document}

\begin{frontmatter}

\title{Transient Behavior of Gossip Opinion Dynamics\\ with Community Structure\thanksref{footnoteinfo}}
\thanks[footnoteinfo]{\emph{Emails:} \texttt{yuxing2@kth.se} (Yu Xing), \texttt{kallej@kth.se} (Karl H. Johansson).}

\author[kth]{Yu Xing}, 
\author[kth]{Karl H. Johansson}

\address[kth]{Division of Decision and Control Systems, School of Electrical Engineering and Computer Science,\\ KTH Royal Institute of Technology, and Digital Futures, Stockholm, Sweden.}

\begin{abstract}
	We study transient behavior of gossip opinion dynamics, in which agents randomly interact pairwise over a weighted graph with two communities. Edges within a community have identical weights different from edge weights between communities. We first derive an upper bound for the second moment of agent opinions. Using this result, we obtain upper bounds for probability that a large proportion of agents have opinions close to average ones. The results imply a phase transition of transient behavior of the process: When edge weights within communities are larger than those between communities and those between regular and stubborn agents, most agents in the same community hold opinions close to the average opinion of that community with large probability, at an early stage of the process. However, if the difference between intra- and inter-community weights is small, most of the agents instead hold opinions close to everyone's average opinion at the early stage. In contrast, when the influence of stubborn agents is large, agent opinions settle quickly to steady state. We then conduct numerical experiments to validate the theoretical results. Different from traditional asymptotic analysis in most opinion dynamics literature, the paper characterizes the influences of stubborn agents and community structure on the initial phase of the opinion evolution.
\end{abstract}



\end{frontmatter}


\section{Introduction}\label{sec_intro}

Opinion dynamics is the study of how personal opinions change through interactions in social networks. Analysis of convergence and stability of opinion dynamics has gained considerable attention in recent decades~\cite{proskurnikov2018tutorial}, but less research has focused on transient behavior of the opinion formation process. Social networks often have topologies where subgroups of nodes are densely connected internally but loosely connected with others (that is, community structure~\cite{fortunato2016community}). Such structure is observed to be able to influence opinion dynamics~\cite{conover2011political,cota2019quantifying}. It is often difficult to determine whether a realistic social network reaches steady state or not, and whether its communities evolve homogeneously at the early stage. So it is necessary to investigate how the opinion formation process, especially during its initial phase, corresponds to its community structure. Such results can also provide insight into design of community detection algorithms based on state observations~\cite{schaub2020blind,xing2023community} and design of model reduction tools for large-scale networks~\cite{cheng2018model
}.


\subsection{Related Work}\label{sec_relatedwork}

Individual opinion can be modeled by  a continuous variable taking values in an interval of real numbers, or a discrete variable in a finite set \cite{proskurnikov2018tutorial,castellano2009statistical}. There are at least three types of continuous-opinion models explaining how interpersonal interactions shape opinion profiles of social networks, namely, models of assimilative, homophily, and negative social influences~\cite{proskurnikov2018tutorial,flache2017models}. Evidences for all three types have been found in recent empirical studies~\cite{de2019learning,friedkin2021group,kozitsin2023opinion}. A crucial example of the first class of models is the DeGroot model~\cite{degroot1974reaching}, in which agents update according to the average of their neighbors' opinions. The Friedkin--Johnsen (FJ) model~\cite{friedkin1990social} generalizes the DeGroot model and allows long-term disagreement, rather than consensus, by assuming that agents are consistently affected by their initial opinions. The Hegselmann--Krause (HK) model~\cite{hegselmann2002opinion
} and the Deffuant--Weisbuch (DW) model \cite{deffuant2000mixing
} are representatives of the second class of models. In these models, agents stay away from those who hold different beliefs, and hence the agents tend to form clusters in the end. Negative influences presence in the third class of models can increase opinion difference between individuals, and make the group end in polarization~\cite{proskurnikov2018tutorial,shi2019dynamics}. 

Most studies of opinion dynamics have focused on asymptotic behavior, attempting to answer why opinion disagreement occurs in the long run even though social interactions tend to reduce opinion difference~\cite{flache2017models,abelson1964mathematical}. In contrast, transient behavior has attracted less attention. As the availability of large-scale datasets from the internet increases, there is a growing need to understand how the process behaves over a finite time interval
~\cite{banisch2012agent,chowell2016mathematical,noorazar2020classical}. Important behavioral dynamics, such as election and online discussion, often have finite duration, and their prediction based on transient evolution is of great interest~\cite{de2019learning,banisch2010empirical}. Extensive amount of information produced by social media nowadays may change public opinions only temporarily~\cite{hill2013quickly}, making it hard for the dynamic process to reach steady state. Asymptotic analysis thus may not be sufficient for understanding such scenarios. Finally, real large-scale networked dynamic processes may converge slowly~\cite{banisch2012agent,lorenz2006consensus
}, but stay close to a certain state for a long time~\cite{barbillon2015network,dietrich2016transient
}. In practice, to distinguish between two types of states can be challenging and requires knowledge of transient system behavior.
The authors in~\cite{banisch2012agent} propose a framework to analyze the transient stage of discrete-state Markov chains that model opinion dynamics. The paper~\cite{barbillon2015network} studies quasi-stationary distributions of a contact process, and~\cite{xiong2017modeling} analyzes the transient opinion profiles of a voter model. The authors in~\cite{dietrich2016transient} provide criteria for detecting transient clusters in a generalized HK model that normally reaches a consensus asymptotically. A recent paper~\cite{s2022finite} studies how opinion difference evolves over finite time intervals for a stochastic bounded confidence model.


The study of community structure can be traced back to~\cite{festinger1949analysis
}, in which a community is defined as a complete subgraph in a network. 
One of the modern definitions for communities is modularity, which characterizes the nonrandomness of a group partition~\cite{newman2004finding}. The stochastic block model (SBM)~\cite{abbe2017community}, generating random graphs with communities, is another popular framework in the literature.
Researchers have studied how community structure of a network influences opinion evolution, based on several models such as the DW model~\cite{gargiulo2010opinion,fennell2021generalized}, the Taylor model
~\cite{baumann2020laplacian}, and the Sznajd model~\cite{si2009opinion}. The paper~\cite{como2016local} studies the DeGroot model with stubborn agents over a weighted graph, and shows that the steady state of agents in the same community concentrates around the state of some stubborn agent.



This paper considers transient behavior of a gossip model with stubborn agents. This model, in which agents randomly interact in pairs, is a stochastic counterpart of the DeGroot model. The model captures the random nature of interpersonal influence and exhibits various behavior. Consensus of the gossip model has been  studied in~\cite{boyd2006randomized,fagnani2008randomized}. The authors in~\cite{acemouglu2013opinion} show that the existence of stubborn agents may explain fluctuation of social opinions, and also that, if the network is highly fluid, then the expected steady state of regular (non-stubborn) agents is close. In contrast, it is shown in~\cite{como2016local} that polarization can emerge, if the model evolves over a weighted graph with two stubborn agents. Studying transient behavior of the gossip model can provide insight into analysis of more complex models, because the linear averaging rule is a key building block of most opinion models. 

\subsection{Contribution}
In this paper, we study transient behavior of the gossip model over a weighted graph with two communities. It is assumed that edges within communities have identical weights different from edge weights between communities. We first obtain an upper bound for the second moment of agent states (Lemma~\ref{lem_second_moment}). Using this bound, we are able to provide probability bounds for agent states concentrating around average opinions (Corollary~\ref{cor_deviationfromTimet}), expected average opinions (Theorem~\ref{thm_deviationFromEt}), and average initial opinions (Theorem~\ref{thm_deviationInterval}). The results reveal a phase transition phenomenon~\cite{holme2006nonequilibrium,biswas2012disorder,shi2016evolution}: When edge weights within communities are larger than those between communities and those between regular and stubborn agents, most agents in the same community have states with small deviation from the average opinion of that community with large probability, at the early stage of the process. But if the difference between intra- and inter-community weights is not so large, most agents in the network have states concentrating around everyone's average opinion with small error and large probability at the early stage (Corollary~\ref{cor_explicitlsd}). In contrast, if weights between regular and stubborn agents are larger than those between regular agents, agent states have a distribution close to their stationary one, right after the beginning of the process (Theorem~\ref{thm_prob3} and Corollary~\ref{cor_prob3}). 


These results indicate that the gossip model has entirely different transient behavior, under different link strength parameters. 
It is known that the expected steady state depends on the positions of stubborn agents~\cite{acemouglu2013opinion}, and the model reaches a consensus if there are no stubborn agents~\cite{boyd2006randomized,fagnani2008randomized}. The obtained results indicate that agents may form transient clusters if the influence of stubborn agents is relatively small. These transient clusters may not be the same as the stationary ones, because they only depend on initial states of regular agents and edge weights between them. 
In addition, the results demonstrate how transient behavior of the model corresponds to community structure, by showing that the difference between intra- and inter-community weights has to be large enough to ensure the existence of such a correspondence and by showing how the relative magnitude of the weights influences the duration of the transient clusters.

The obtained results can be directly applied to community detection based on state observations~\cite{schaub2020blind,xing2023community}. Suppose that a network is unknown but several snapshots of an opinion dynamic are available. Then our results ensure that partitioning agent states can recover the community labels of agents, if the states are collected in a transient time interval and intra-community influence is large.
The results can also provide insight into predicting and distinguishing transient behavior of an opinion formation process in practice~\cite{banisch2012agent,banisch2010empirical}. 
For example, when the size of a network and the difference between intra- and inter-community interaction strength are large, the duration of transient behavior is expected to be large as well. Given an estimate of the runtime of a process, we may determine whether the current clusters are steady or not.  Finally, exploiting properties of transient clusters can help improve model reduction at the initial phase of dynamics over large-scale networks with community structure~\cite{cheng2018model}. If agents in the same community have states close to each other at the early stage of the process, then we may track the opinion formation process with much less parameters than the original system, by replacing topological data with community labels.

In the early work~\cite{xing2022what}, we study how the expectation of agent states concentrates around average initial states, which follows from Lemma~\ref{lem_expression_expectation_xt} of the current paper. Here we directly investigate how agent states evolve by analyzing the second moment and thus provide more detailed characterization of transient behavior.



\subsection{Outline}
In Section~\ref{sec_problem} we introduce the model and the problem studied in the paper. Section~\ref{sec_results} provides theoretical results. Numerical experiments are presented in Section~\ref{sec_simulation}. Section~\ref{sec_conclusion} concludes the paper. Some proofs are given in the Appendix.\\ 

\noindent\textbf{Notation.}
Denote the $n$-dimensional Euclidean space by $\mathbb{R}^n$, the set of $n\times m$ real matrices by $\mathbb{R}^{n\times m}$, the set of nonnegative integers by $\mathbb{N}$, and $\mathbb{N}^+ = \mathbb{N}\setminus\{0\}$. Denote the natural logarithm by $\log x$, $x\in \RR$. Let $\mathbf{1}_n$ be the all-one vector with dimension $n$, $e_i^{(n)}$ be the $n$-dimensional unit vector with $i$-th entry being one, $I_n$ be the $n\times n$ identity matrix, and $\bfo_{m,n}$ be the $m\times n$ all-zero matrix. Denote  the Euclidean norm of a vector by $\|\cdot\|$. 
For a vector $x\in \mathbb{R}^n$, denote its $i$-th entry by $x_i$, and for a matrix $A \in \mathbb{R}^{n\times n}$, denote its $(i,j)$-th entry by $a_{ij}$ or $[A]_{ij}$. 
The cardinality of a set $\mtcs$ is denoted by $|\mtcs|$. The function $\mathbb{I}_{[\textup{property}]}$ is the indicator function equal to one if the property in the bracket holds, and equal to zero otherwise.  Denote the expectation of a random vector $X$ by $\mathbb{E}\{X\}$.  For two sequences of real numbers, $f(n)$ and $g(n) > 0$, $n\in\NN$, we write $f(n) = O(g(n))$ if $|f(n)| \le C g(n)$ for all $n\in \NN$ and some $C > 0$, and $f(n) = o(g(n))$ if $|f(n)|/g(n) \to 0$. Further assuming $f(n)$ to be nonnegative, we say that $f(n) = \omega(g(n))$ if $g(n) = o(f(n))$, that $f(n) = \Omega(g(n))$ if there is $C>0$ such that $f(n) \ge C g(n)$ for all $n \in \NN$, and that $f(n) = \Theta(g(n))$ if both $f(n) = O(g(n))$ and $f(n) = \Omega(g(n))$ hold. For $x,y \in \RR$, denote $x\vee y \Let\max\{x,y\}$ and $x\wedge y\Let \min\{x,y\}$.

\section{Problem Formulation}\label{sec_problem}

The gossip model with stubborn agents is a random process evolving over an undirected graph $\mathcal{G} = (\mathcal{V}, \mathcal{E},A)$, where $\mtcv$ is the node set with $|\mathcal{V}| = n \ge 2$, $\mtce$ is the edge set, and $A = [a_{ij}] \in \RR^{n\times n}$ is the weighted adjacency matrix.
The graph $\mathcal{G}$ has no self-loops (i.e., $a_{ii} = 0$, $\forall i \in \mtcv$). $\mathcal{V}$ contains two types of agents, regular and stubborn, denoted by $\mathcal{V}_r$ and $\mathcal{V}_s$, respectively (so $\mathcal{V} = \mathcal{V}_r \cup \mathcal{V}_s$ and $\mathcal{V}_r \cap \mathcal{V}_s = \emptyset$). 
In this paper we assume that the regular agents form two disjoint communities $\mtcv_{r1}$ and $\mtcv_{r2}$, and denote $\mtcc_i = k$ if $i \in \mtcv_{rk}$, $k =1,2$. We call $\mtcc$ the community structure of the graph.
Regular agent~$i$ has state $X_i(t) \in \mathbb{R}$ at time $t \in \mathbb{N}$, and stubborn agent~$j$ has state $\bfz^s_j$. Stacking these states, we denote the state vector of regular agents at time $t$ by $X(t) \in \mathbb{R}^{n_r}$ and the state vector of stubborn agents by $\bfz^s \in \RR^{n_s}$, where $n_r \Let |\mtcv_r|$ and $n_s \Let |\mtcv_s|$.
The random interaction of the gossip model is captured by an interaction probability matrix $W = [w_{ij}] \in \mathbb{R}^{n\times n}$ satisfying that $w_{ij} = w_{ji} = a_{ij}/\alpha$, where $\alpha = \sum_{i=1}^n\sum_{j=i+1}^n a_{ij}$ is the sum of all edge weights. Hence $\mathbf{1}^T W \mathbf{1}/2 = 1$.
 At time $t$, edge $\{i,j\}$ is selected with probability $w_{ij}$ independently of previous updates, and agents update as follows, 
 \begin{align}\label{eq_update_compact_regular}
    X(t+1) = Q(t) X(t) + R(t) z^s,
\end{align}
where $Q(t) \in \RR^{n_r\times n_r}$, $R(t) \in \RR^{n_r\times n_s}$, and
\begin{align*}
	&[Q(t),~R(t)] = \\&\begin{cases}
    [I_{n_r} - \frac12 (e_i^{(n_r)} - e_j^{(n_r)} )(e_i^{(n_r)}  - e_j^{(n_r)} )^T,~~~\bfo_{n_r,n_s}], \\
    	\qquad\qquad\qquad\qquad\qquad\qquad\qquad\qquad\qquad \text{if } i, j \in \mathcal{V}_r,\\
    [I_{n_r} - \frac12 e_i^{(n_r)}(e_i^{(n_r)})^T,~~~\frac12 \bfe_i^{(n_r)} (e_j^{(n_s)})^T], \\
    	\qquad\qquad\qquad\qquad\qquad\qquad\qquad\quad~~~  \text{if } i \in \mathcal{V}_r,  j \in \mathcal{V}_s.
    \end{cases}
\end{align*} 
That is, only regular agents in $\{i,j\}$ update their states to the average of the selected agents' previous states.

We study how community structure and stubborn agents influence transient behavior of agent states $X(t)$. The community structure will be defined in Section~\ref{sec_mainres}. By transient behavior we mean a property of $X(t)$ that holds over a finite time interval, as opposed to asymptotic behavior that holds as time $t\to \infty$. 
As mentioned in Section~\ref{sec_relatedwork},  agents in the same community tend to have similar states, but how well and how long these clusters form still require rigorous analysis. We characterize the transient clusters of $X(t)$ based on three types of references: (1) agents' average states within communities and everyone's average state at time $t$, (2) expected average states at time $t$, and (3) average initial states. As a comparison, we also study the time when the distribution of $X(t)$ is close to the stationary distribution. To sum up, the considered problem is as follows.


\begin{figure*}[t]
	\centering
	\includegraphics[scale=0.35]{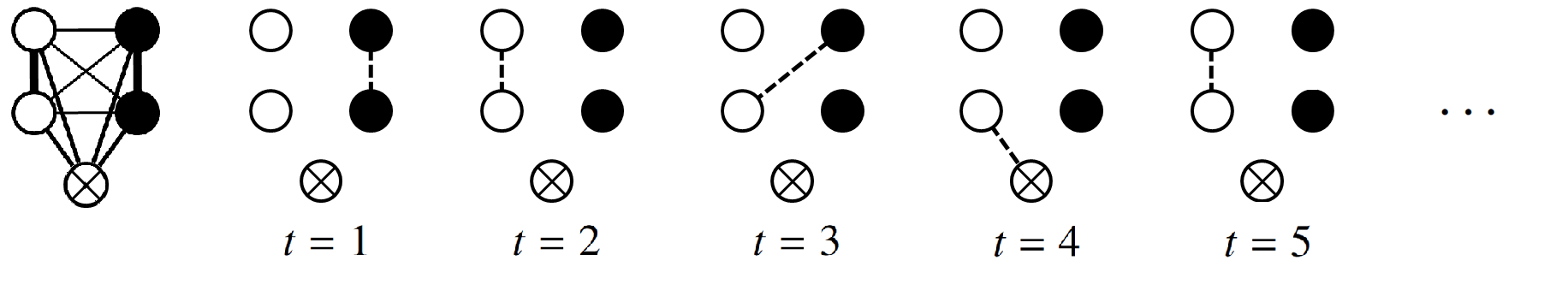}
	\caption{\label{fig_illus_asmp} Illustration of Assumption~\ref{asmp_1}. The graph on the left demonstrates the underlying network with two communities (dots and circles) and one stubborn agent (the circle with a cross). Solid lines represent weighted edges. The weights are indicated by line thickness. Edge weights within communities are larger than between communities ($l_s^{(r)} > l_d^{(r)}$). The edge weights between regular agents and the stubborn agent are the same.  The rest of the graphs show random interactions between agents, represented by dashed lines. Agents interact more often if they have an edge with a larger weight.}	
\end{figure*}

\textbf{Problem.} Given the initial states $X(0)$, the stubborn states $\bfz^s$, the community structure $\mtcc$, and the weighted adjacency matrix $A$, provide bounds for the deviation of $X(t)$ from the three types of average states over finite time intervals, and bounds for the time when the distribution of $X(t)$ is close to the stationary distribution.

Probability bounds for the deviation of $X(t)$ from average states at time $t$ are given in Corollary~\ref{cor_deviationfromTimet}, which is a consequence of second-moment analysis (Lemma~\ref{lem_second_moment}). The deviation from expected average states is analyzed in Theorem~\ref{thm_deviationFromEt}. This result is stronger than the previous one because the references are deterministic. Theorem~\ref{thm_deviationInterval} shows concentration of $X(t)$ around average initial states. Theorem~\ref{thm_prob3} gives a lower bound of the time when $X(t)$ is close to steady state.

\section{Theoretical Analysis}\label{sec_results}
We study transient behavior of gossip model with two communities. The analysis provides crucial insight into understanding transient behavior of the model under general conditions. Main results are presented in Section~\ref{sec_mainres} and a discussion is given in Section~\ref{sec_dis}.

\subsection{Main Results}\label{sec_mainres}

We assume that the regular agents form two disjoint communities $\mtcv_{r1}$ and $\mtcv_{r2}$ with equal size. For simplicity, sort the agents as follows: $\mtcv_{r1} = \{1,\dots, r_0n/2\}$, $\mtcv_{r2} = \{1 + r_0n/2,\dots, r_0n\}$, and $\mtcv_s = \{1+r_0n, \dots, n\}$, with $r_0 \in (0,1)$ such that $r_0n = n_r$ is an even integer. The proportion of stubborn agents is denoted by $s_0 := 1 - r_0$. 
We introduce the following assumptions, illustrated in Fig.~\ref{fig_illus_asmp}, for the weighted adjacency matrix $A$ of graph $\mtcg$. 

\begin{asmp}[Network topology]\label{asmp_1}~
	\begin{enumerate}[(i)]
		\item There exist $l_s^{(r)}, l_d^{(r)} \in (0,1)$, depending on $n$, such that $a_{ij} = l_s^{(r)} = l_s^{(r)}(n)$ for $i,j\in \mtcv_r$ with $i\not=j$ and $\mtcc_i = \mtcc_j$, $a_{ij} = l_d^{(r)} = l_d^{(r)}(n)$ for $i,j\in \mtcv_r$ with $\mtcc_i \not= \mtcc_j$.
		\item There exist $ l_{ij}^{(s)} \in [0,1)$ with $1\le i \le r_0n$ and $1\le j\le s_0n$, depending on $n$,  such that $a_{i, r_0n+j} = a_{r_0n+j, i} = l_{ij}^{(s)} = l_{ij}^{(s)}(n)$. For $r_0n+1 \le i,j\le n$, $a_{ij} = 0$.
		\item There exists a nonnegative number $l^{(s)}$ depending on $n$ such that $\sum_{1\le j \le s_0n} l_{ij}^{(s)} = l^{(s)} = l^{(s)}(n)$ for all $i \in \mtcv_r$. 
	\end{enumerate}
\end{asmp}


\begin{rmk}\label{rmk_asmp1}
    Assumption~\ref{asmp_1}~(i) indicates that the graph on regular agents is a weighted graph such that edges between agents in the same community have the same weight $l_s^{(r)}$ whereas those between communities have weight $l_d^{(r)}$. In other words, the influence strength between agents depends on their community labels. Such a weighted graph can be treated as the expected adjacency matrix of an SBM, in which nodes are assigned with community labels, and an edge exists with probability depending on the community labels of that edge's endpoints. 
    We introduce this simplified assumption to highlight the phase transition phenomenon in the transient phase of the dynamics, and proofs under this assumption are still nontrivial. It is possible to generalize the results to the SBM case by considering the concentration of adjacency matrices~\cite{chung2011spectra}. Although assuming that interpersonal influence depends only on community labels is a simplified setting, this model has been found to be efficient also in empirical studies~\cite{de2019learning}. 
    Note that the adjacency matrix~$A$ has zeros entries in the diagonal from the assumption that the graph has no self-loops.
	Parameter $l^{(s)}$ given in~Assumption~\ref{asmp_1}~(iii) is the sum of edge weights between one regular agent and all stubborn agents, and thus represents the total influence of stubborn agents on this regular agent. We assume that this sum is the same for all regular agents for analysis simplicity. The results can be extended to the case where the weight sums have upper and lower bounds. 
\end{rmk}

We further impose an assumption for the initial vector.

\begin{asmp}[Initial condition]\label{asmp_2}
	The initial regular states $X(0)$ and the stubborn states $\bfz^s$ are deterministic, and satisfy that $|X_i(0)| \le c_x$ and $|\bfz^s_j| \le c_x$, for all $1\le i \le r_0n$ and $1\le j\le s_0n$, and some  $c_x>0$.
\end{asmp}

From the definitions of $Q(t)$ and $R(t)$, it follows that
\begin{align}\nonumber
	&\bar{Q} \Let \EE\{Q(t)\} \\\label{eq_barQ}
	& = I - \frac{1}{2\alpha}
	\begin{bmatrix}
	d_1 & -a_{12} & \cdots & -a_{1,r_0n}\\
	-a_{21} & \ddots & & \vdots\\
	\vdots &  & & \vdots \\
	-a_{r_0n,1} & \cdots & -a_{r_0n,r_0n-1} & d_{r_0n}
	\end{bmatrix}, \\\nonumber
	&\bar{R} \Let \EE\{R(t)\} \\\label{eq_barR}
	&= \frac{1}{2\alpha} \tilde{M} \Let \frac{1}{2\alpha} 
	\begin{bmatrix}
		a_{1,r_0n+1} & \cdots & a_{1, n}\\
		\vdots & & \vdots\\
		a_{r_0n, r_0n+1} & \cdots & a_{r_0n,n}
	\end{bmatrix},
\end{align}
where $d_i = \sum_{j\in \mtcv} a_{ij}$, $i \in \mtcv_r$. Note that Assumption~\ref{asmp_1} implies $d_i = r_0n(l_s^{(r)} + l_d^{(r)})/2 + l^{(s)} - l_s^{(r)} \teL \bar{d}$, $i\in \mtcv_r$,  so
\begin{align*}
    \bar{Q} = I - \frac{1}{2\alpha} [(\bar{d}+l_s^{(r)})I - \tilde{A}],
\end{align*}
where 
\begin{align*}
    \tilde{A} \Let \begin{bmatrix}
        \bfl_{r_0n/2} & \bfo_{r_0n/2} \\
        \bfo_{r_0n/2} & \bfl_{r_0n/2}
    \end{bmatrix} \begin{bmatrix}
        l_s^{(r)} & l_d^{(r)} \\
        l_d^{(r)} & l_s^{(r)}
    \end{bmatrix} 
    \begin{bmatrix}
        \bfl_{r_0n/2} & \bfo_{r_0n/2} \\
        \bfo_{r_0n/2} & \bfl_{r_0n/2}
    \end{bmatrix}^T.
\end{align*}
It can be shown that $\tilde{A}$ has a simple eigenvalue $r_0n(l_s^{(r)} + l_d^{(r)})/2$, and a simple eigenvalue $r_0n(l_s^{(r)} - l_d^{(r)})/2$, and the corresponding unit vectors are $\eta \Let \bfl_{r_0n}/\sqrt{r_0n}$ and $\xi := [\bfl_{r_0n/2}^T~-\bfl_{r_0n/2}^T]/\sqrt{r_0n}$, respectively. Since $\tilde{A}$ is symmetric, it has the eigenvalue zero with multiplicity $r_0n-2$ with orthogonal unit eigenvalues $w^{(i)}$, $3\le i \le r_0n$.  Moreover, $\eta$, $\xi$, $w^{(3)}$, $\dots$, $w^{(r_0n)}$ form an orthonormal basis of $\RR^{r_0n}$. Denoting $\lambda_1 \Let  l^{(s)}/(2\alpha)$, $\lambda_2 \Let (l_d^{(r)} r_0n + l^{(s)})/(2\alpha)$, and $\lambda_3 \Let [(l_s^{(r)} + l_d^{(r)}) r_0n/2 + l^{(s)}]/(2\alpha)$, we present the following summary of properties of $\bar{Q}$ and orthogonal vectors, which will be used later.
\begin{lemm}\label{lem_eigen}
    Under Assumption~\ref{asmp_1}, the following hold.\\
    (i) The matrix $\bar{Q} \in \RR^{r_0n\times r_0n}$ has a simple eigenvalue $1-\lambda_1$ with a unit eigenvector $\eta$, a simple eigenvalue $1 - \lambda_2$ with a unit eigenvector $\xi$, and an eigenvalue $1 - \lambda_3$  with multiplicity $r_0n-2$ and with unit eigenvectors $w^{(i)}$, $3\le i \le r_0n$. In addition, the vectors $\eta$, $\xi$, $w^{(3)}$, $\dots$, $w^{(r_0n)}$ form an orthonormal basis of $\RR^{r_0n}$.\\
    (ii) If $\{x^{(i)} \in \RR^n, 1\le i \le n\}$ is an orthonormal basis of $\RR^n$, then it holds for all $z \in \RR^n$ and $1\le j \le n$ that 
    \begin{align*}
        I_n &= \sum_{i=1}^n x^{(i)} (x^{(i)})^T,\\
        \|z\|^2 &= \Big\|\sum_{i=1}^n x^{(i)} (x^{(i)})^T z \Big\|^2 = \sum_{i=1}^n \|x^{(i)} (x^{(i)})^T z\|^2 \\
        &= \Big\| \sum_{i=1}^j x^{(i)} (x^{(i)})^T z \Big\|^2 + \Big\|\sum_{i=j+1}^n x^{(i)} (x^{(i)})^T z \Big\|^2.
    \end{align*}
\end{lemm}
Before presenting main theorems, we provide several properties of the gossip model. The first lemma concerns the explicit expression of the weight sum  of the graph $\mtcg$.

\begin{lemm}\label{lem_alpha_value}
	Under Assumption~\ref{asmp_1}, the weight sum $\alpha = r_0n[(l_s^{(r)} + l_d^{(r)}) r_0n + 4 l^{(s)} - 2l_s^{(r)}]/4$.
\end{lemm}

\begin{pf}
	The conclusion follows directly from the definition of $\alpha$ and Assumption~\ref{asmp_1}. \hfill$\Box$
\end{pf}

The next lemma gives the expression of $\EE\{X(t)\}$, and shows how $\EE\{X(t)\}$ evolves by decomposing it into three parts, which correspond to the eigenspaces of $\bar{Q}$.

\begin{lemm}\label{lem_expression_expectation_xt}
	Suppose that Assumption~\ref{asmp_1} holds. Then the expectation of $X(t)$ satisfies that, for all $t \in \NN$,
	\begin{align*}
	&\EE\{X(t)\}\\
	 &= (1 - \lambda_1)^t \eta \eta^T X(0) + \frac{1}{\lambda_1} [1 - (1 - \lambda_1)^t] \eta \eta^T \bar{R} \bfz^s \\
	& + (1 - \lambda_2)^t \xi \xi^T X(0) + \frac{1}{\lambda_2} [1 - (1 - \lambda_2)^t] \xi \xi^T \bar{R} \bfz^s \\
	& + (1 - \lambda_3)^t \sum_{i = 3}^{r_0n} w^{(i)} (w^{(i)})^TX(0)\\
	& + \frac{1}{\lambda_3} [1 - (1 - \lambda_3)^t] \sum_{i = 3}^{r_0n} w^{(i)} (w^{(i)})^T\bar{R} \bfz^s.
	\end{align*}
\end{lemm}

\begin{pf}
	 Lemma~\ref{lem_eigen} yields that
\begin{align}\nonumber
	\bar{Q} &= (1 - \lambda_1) \eta \eta^T +  ( 1 - \lambda_2) \xi \xi^T  \\\label{eq_Q_decomp}
	&+ ( 1 - \lambda_3) \sum_{i = 3}^{r_0n} w^{(i)} (w^{(i)})^T.
\end{align}
Then the result follows from~\eqref{eq_update_compact_regular}--\eqref{eq_barR}.
\hfill$\Box$
\end{pf}

We further introduce the following technical assumption.
\begin{asmp}\label{asmp_3} 
	Denote $\tilde{l}^{(s)}_+ \Let \max_{1\le j \le s_0n}\sum_{i\in \mtcv_r} l_{ij}^{(s)}$. 
    It holds that $\tilde{l}^{(s)}_+ \le c_l l^{(s)}$ 
    for some  constant $c_l > 0$. \\
\end{asmp}

\begin{rmk}
    The assumption indicates that the maximum influence strength of a stubborn agent on all regular agents is of the same order of $l^{(s)}$.
    Otherwise, the influence of stubborn agents may not be homogeneous, which could be hard to analyze. 
\end{rmk}

Now we present a lemma that will be used in the proof of main theorems.
Denote $X^{\eta}(t) := \eta\eta^TX(t)$, $X^{\xi}(t) := \xi\xi^TX(t)$, $\Gamma := \sum_{i = 3}^{r_0n}  w^{(i)} (w^{(i)})^T$ (thus, $\Gamma^T \Gamma = \Gamma^2 = \Gamma$), and $X^{\Gamma}(t) := \Gamma X(t)$. 
From Lemma~\ref{lem_eigen},
$\eta \eta^T + \xi \xi^T + \Gamma  = I_{r_0n}$, and we have the decomposition $X(t) = X^{\eta}(t) + X^{\xi}(t) + X^{\Gamma}(t)$. Further, let $X^{\bot}(t) \Let \tilde{\Gamma} X(t)$, where $\tilde{\Gamma} = \xi\xi^T + \Gamma$. We get another decomposition of $X(t)$: $X(t) = X^{\eta}(t) + X^{\bot}(t)$ and $X^{\eta}(t)^T X^{\bot}(t) = 0$. 
Note that $X^\eta(t)+X^\xi(t)$ and $X^\eta(t)$ represent the average states in each community and everyone's average state, respectively (the $i$-th entry of $X^\eta(t)+X^\xi(t)$ is $(2\sum_{j \in \mtcv_{r\mtcc_i}} X_j(t))/(r_0n)$). So the two decompositions reveal dynamics of average states.
The following lemma gives two upper bounds for the second moment of $X(t)$ under the two decompositions.
\begin{lemm}\label{lem_second_moment}
	Suppose that Assumptions~\ref{asmp_1}--\ref{asmp_3} hold. Denote $c_s \Let \sqrt{s_0/r_0}$. It holds for $t\in \NN$ and $n \ge 4/r_0$ that
	\begin{align}\nonumber
		&~\EE\{\|X(t)\|^2\} \\\nonumber
		&\le  (1-\lambda_1)^t \|X^\eta(0)\|^2 + (1-\lambda_2)^t \|X^\xi(0)\|^2 + (1-\lambda_3)^t \\\nonumber
        &~\|X^\Gamma(0)\|^2 + (1 \wedge \lambda_1 t) C_{11} c_x^2 r_0n  + \Big(\frac{\lambda_2}{\lambda_3} \wedge \lambda_2t \Big) [(1-\lambda_2)^t \\\label{eq_2ndmomentbound_local}
        &~ \|X^\xi(0)\|^2 + c_x^2] + (1 \wedge \lambda_2 t) c_x^2,\\\nonumber
        &~\EE\{\|X(t)\|^2\} \\ \nonumber
        &\le (1-\lambda_1)^t \|X^{\eta}(0)\|^2 + (1-\lambda_2\wedge\lambda_3)^{t} \|X^{\bot}(0)\|^2 \\ \label{eq_2ndmomentbound_global}
        & + (1\wedge \lambda_1 t) C_{21} c_x^2 r_0n   + \Big(\frac{\lambda_1}{\lambda_2 \wedge \lambda_3} \wedge \lambda_1t\Big) C_{22} c_x^2 r_0n ,
    \end{align}
    where $C_{11} \Let 3 + c_l/2 + 16 c_s c_l + (3c_l + 23)/(2 r_0n)$, $C_{21} \Let 4c_sc_l + (5+c_l)/(2r_0n) $, and $C_{22} \Let 3 +  c_l/2 + 4 c_s c_l+  2/(r_0n)$.
\end{lemm}
\vspace{-3ex}
\begin{pf}
    The main idea of the proof is to separately bound the terms $\EE\{\|X^\eta(t)\|^2\}$, $\EE\{\|X^\xi(t)\|^2\}$, $\EE\{\|X^\Gamma(t)\|^2\}$, and $\EE\{\|X^{\bot}(t)\|\}^2$. The conclusions then follow from Lemma~\ref{lem_eigen}. See Appendix~\ref{sec_append_proof_lem3} for the details. \hfill$\Box$
\end{pf}
\begin{rmk}
    Lemma~\ref{lem_second_moment} provides two bounds for how the second moment of agent states evolve over time. Three types of terms appear in the bounds. The first type is an exponentially decreasing term $(1-\lambda_i)^t$, indicating the rate of the averaging update. The second one is a linearly increasing term $1 \wedge \lambda_i t$, showing cumulative influence of stubborn agents. Lastly, the ratios $\lambda_2 t \wedge (\lambda_2/\lambda_3)$ and $\lambda_1t \wedge [\lambda_1/(\lambda_2 \wedge\lambda_3)]$ indicate the effect of relative influence strength between regular and stubborn agents. The bounds become trivial for $t = \omega(1/\lambda_1)$, and thus only describe transient behavior.
\end{rmk}

\begin{figure*}[t]
	\centering
	\subfigure[\label{fig_illus_case1a}The case where $l_s^{(r)} = (\log^{2.5} n)/n$, $l_d^{(r)} = (\log  n)/n$, and $l^{(s)} = (s_0 \log n)/2$. Most agent states in the same community are close to the expected average state of that community (black dashed lines).]{\qquad~~~ \includegraphics[scale=0.32]{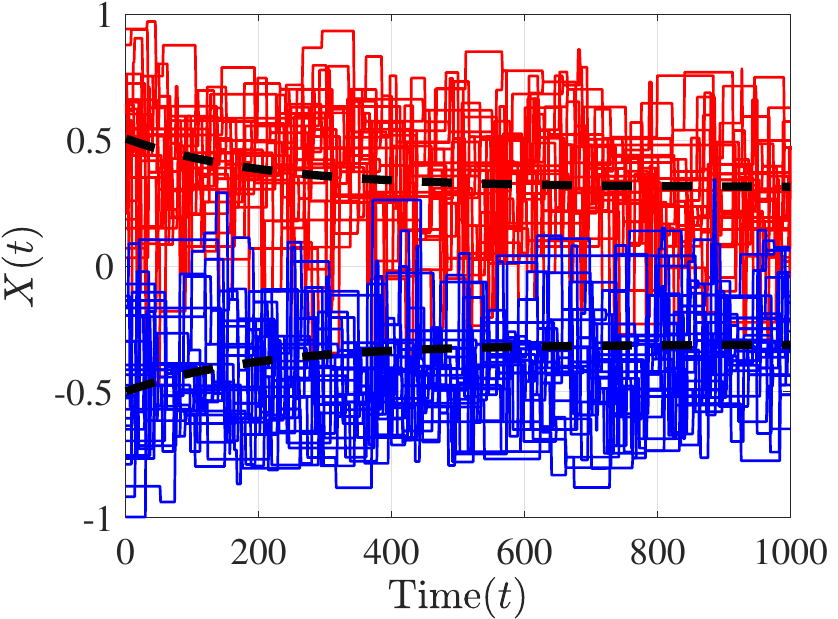} \qquad~~~ }\quad
	\subfigure[\label{fig_illus_case1b}The case where $l_s^{(r)} = (\log^{2.5} n)/n$, $l_d^{(r)} = (\log  n)/n$, and $l^{(s)} = (s_0 \log^2 n)/2$. Agents behave similar to Fig.~\ref{fig_illus_case1a}, but move towards the origin in the long run.]{\qquad~~~ \includegraphics[scale=0.32]{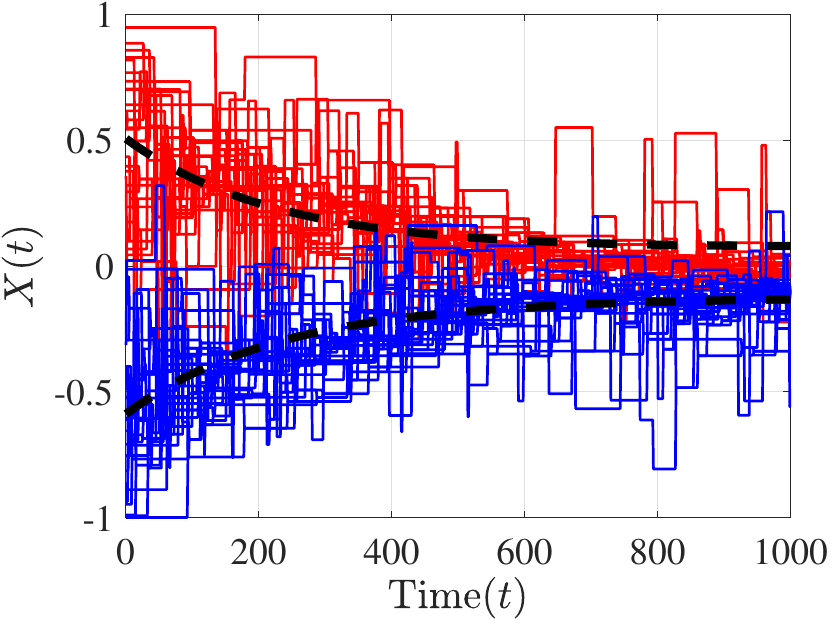} \qquad~~~}
	\subfigure[\label{fig_illus_case2}The case where $l_s^{(r)} = (\log^{2.5} n)/n$, $l_d^{(r)} = (\log^{2.4}  n)/n$, and $l^{(s)} = (s_0 \log n)/2$. Most states are close to everyone's expected average state (the black dashed line), and keep fluctuating
    .]{\qquad~~~ \includegraphics[scale=0.32]{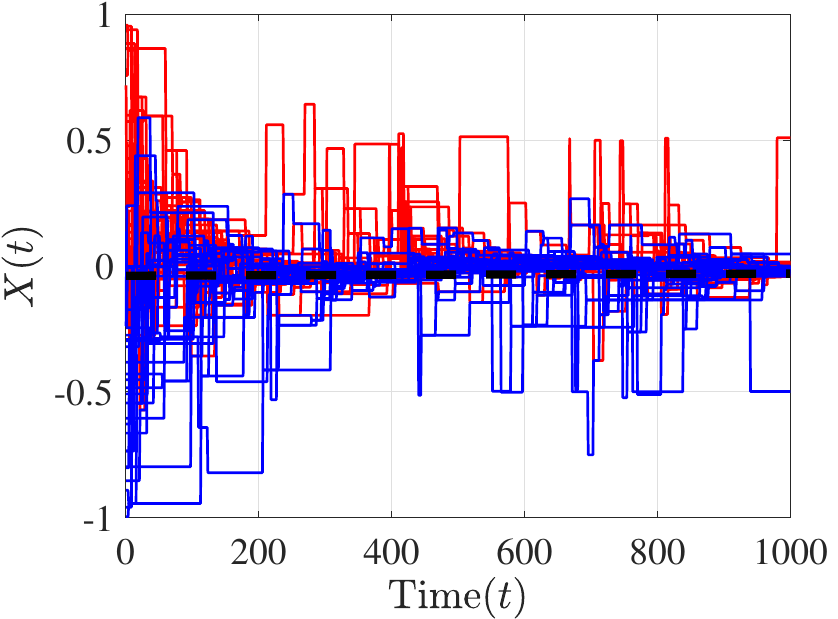} \qquad~~~}\quad
	\subfigure[\label{fig_illus_case3}The case where $l_s^{(r)} = (\log n)/n$, $l_d^{(r)} = (\log n)/n$, and $l^{(s)} = (s_0 \log^3 n)/2$. Most agents move quickly towards stubborn agents during the initial phase, and stay polarized
    .]{\qquad~~~ \includegraphics[scale=0.32]{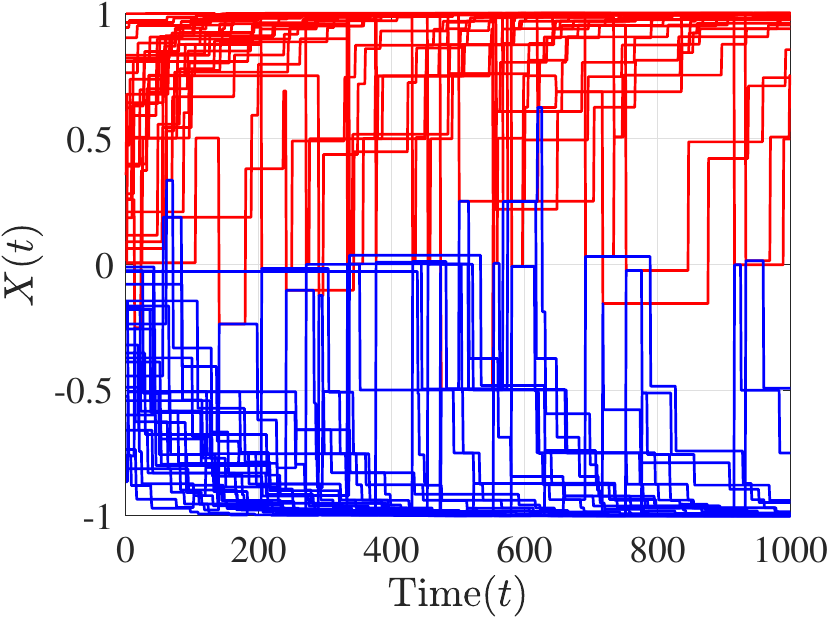} \qquad~~~}
	\caption{\label{fig_illus} Behavior of the gossip model. The figure illustrates the evolution of regular-agent states under four sets of parameters. The color of a trajectory represents the community label of that agent (red represents $\mtcv_{r1}$, and blue represents $\mtcv_{r2}$).}
\end{figure*}

\begin{figure*}[t]
	\centering
        \subfigure[\label{fig_expcase1}Concentration around expected average states within communities, where $n=60$, $l_s^{(r)} = (\log^{2.5} n)/n$, $l_d^{(r)} = (\log  n)/n$, and $\eps = 0.3$.]{\qquad~~~\includegraphics[scale=0.34]{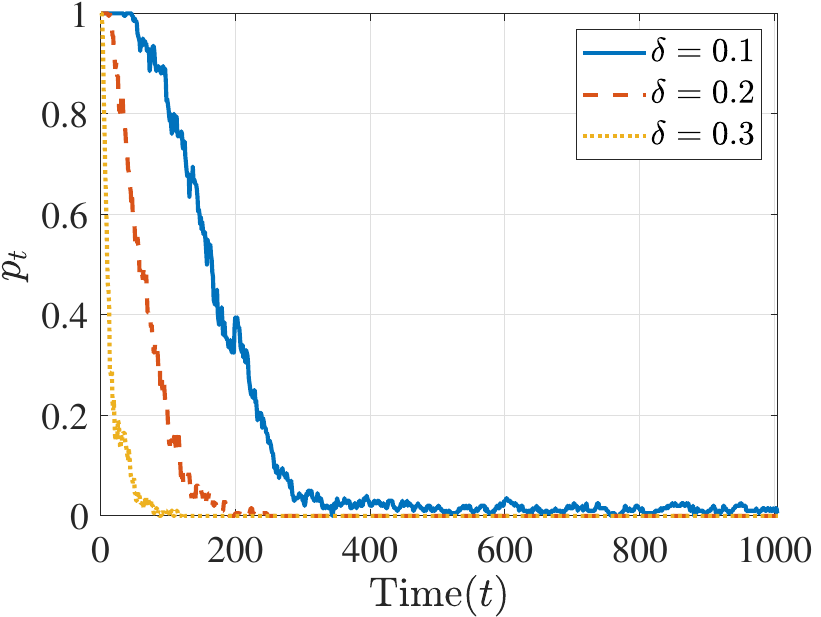}\qquad~~~}~~
	\subfigure[\label{fig_expcase2}Concentration around everyone's expected average state, where $n=60$, $l_s^{(r)} = (\log^{2.5} n)/n$, $l_d^{(r)} = (\log^{2.4}  n)/n$, and $\eps = 0.2$.]{\qquad~~~\includegraphics[scale=0.34]{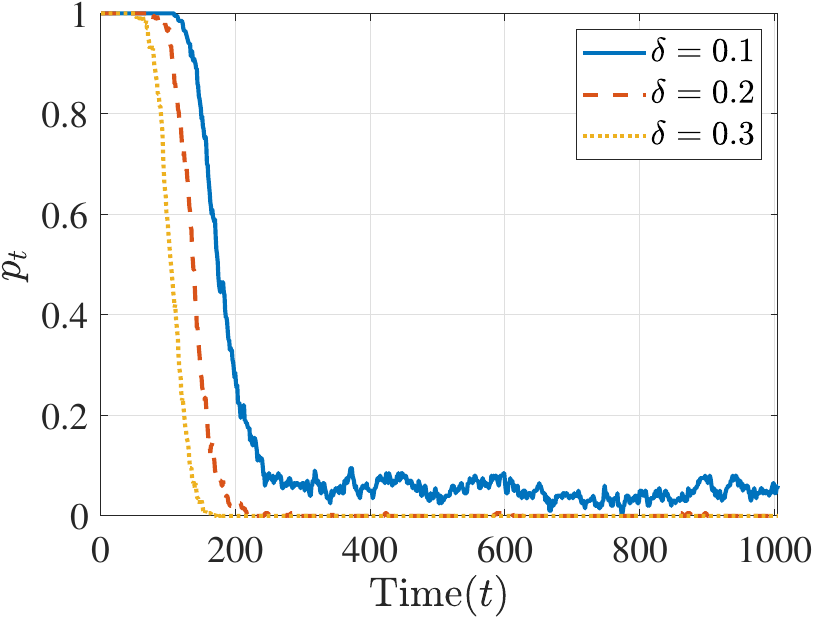}\qquad~~~}\\
        \subfigure[\label{fig_initcase1_60}Concentration around average initial states within communities, where $n$, $l_s^{(r)}$, $l_d^{(r)}$, and $\eps$ are the same as~(a).]{~~~\includegraphics[scale=0.34]{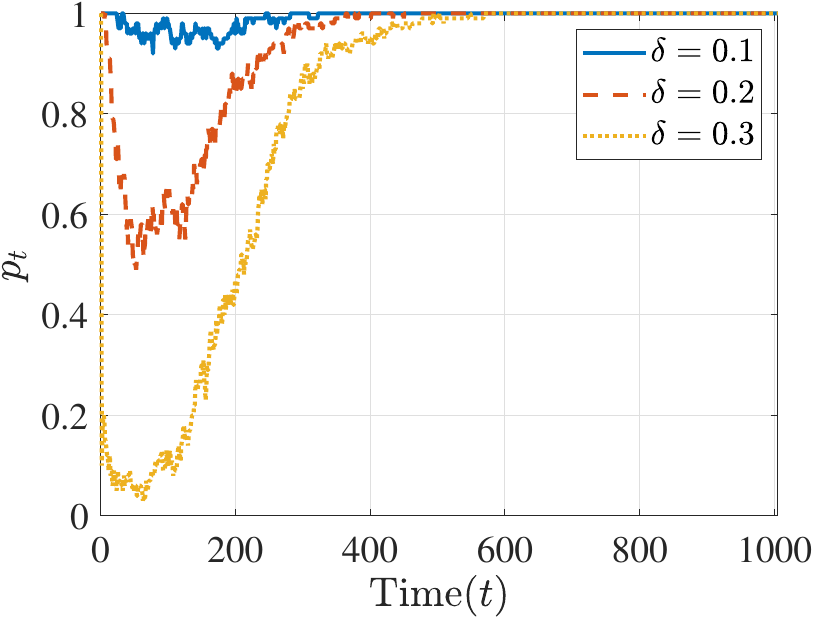}~~~}~~
        \subfigure[\label{fig_initcase1_500}Concentration around average initial states within communities, where $l_s^{(r)}$, $l_d^{(r)}$, and $\eps$ are the same as (a) and $n=500$.]{~~~~~~~\includegraphics[scale=0.34]{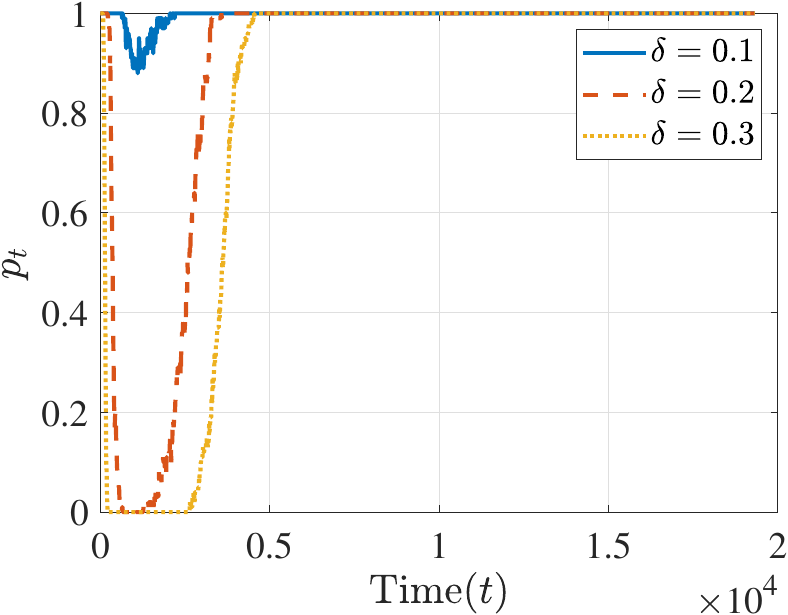}~~~~~~~}~~
        \subfigure[\label{fig_initcase2}Concentration around everyone's average initial state, where $n$, $l_s^{(r)}$, $l_d^{(r)}$, and $\eps$ are the same as (b).]{~~~\includegraphics[scale=0.34]{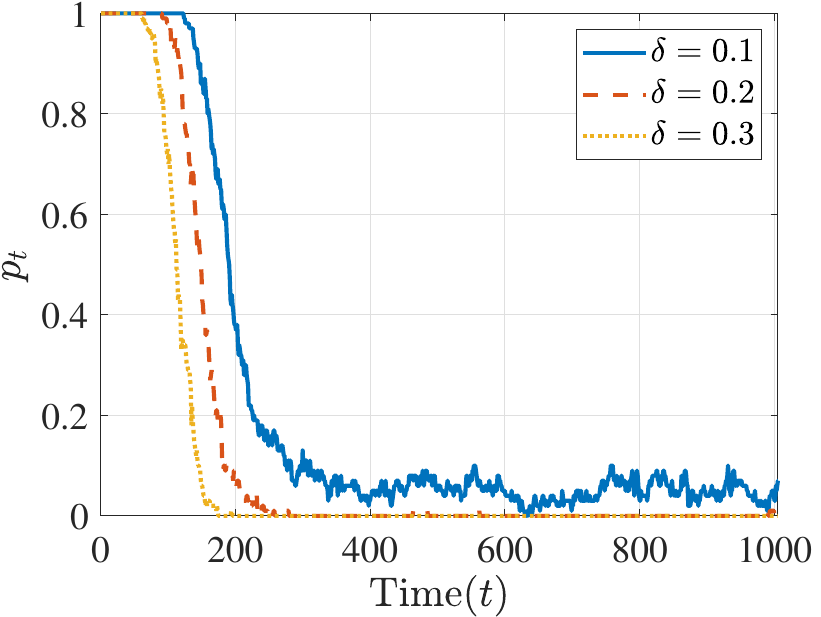}~~~}
	\caption{\label{fig_cases} The probability $p_t$ of at least $\delta$ proportion of agents with states at least $\eps c_x$ away from average states at time $t$. In all experiments, $l^{(s)} = (s_0 \log n)/2$. The expected average states are considered in (a--b), whereas the average initial states in (c--e).}
\end{figure*}

\vspace{-1ex}
An immediate consequence of the preceding analysis is that the difference between the agent states and average states at each time step can be bounded. 
For two vectors $X, Y \in \RR^{r_0n}$ and $\eps \in (0,1)$, we denote the set of agents~$i$ such that $|X_i - Y_i| > \eps c_x$ (i.e., the difference between $X_i$ and $Y_i$ is large) by 
\begin{align*}
    \mtcs(X,Y,\eps) \Let \{i\in \mtcv_r: |X_i - Y_i| > \eps c_x \}.
\end{align*}
Then the proof of Lemma~\ref{lem_second_moment} ensures the following result.
\begin{corr}\label{cor_deviationfromTimet}
Suppose that Assumptions~\ref{asmp_1}--\ref{asmp_3} hold. Then it holds for $t\in \NN$, $n \ge 4/r_0$, and $\eps,\delta\in(0,1)$ that 
    \begin{align}\nonumber
        &~\PP\{|\mtcs(X(t),X^\eta(t)+X^\xi(t),\eps)| \ge \delta r_0n\} \\ \nonumber
        &\le \frac{1}{\eps^2\delta} \Big[ (1-\lambda_3)^t \frac{\|X^\Gamma(0)\|^2}{c_x^2r_0n} + \Big(\frac{\lambda_1}{\lambda_3} \wedge \lambda_1t\Big) \Big( 3 + \frac{c_l}{2}\\ 
        \nonumber
        & + 10 c_sc_l + \frac{c_l + 13}{2r_0n}\Big) + \Big(\frac{\lambda_2}{\lambda_3} + \lambda_2t\Big) \Big( \frac{(1-\lambda_2)^t\|X^\xi(0)\|^2}{c_x^2r_0n} \\ \label{eq_Xgammat_probbound}
        &+ \frac{1}{r_0n}\Big) \Big],\\ \nonumber
        &~\PP\{|\mtcs(X(t),X^\eta(t),\eps)| \ge \delta r_0n\} \\ \nonumber
        &\le \frac{1}{\eps^2\delta} \Big[ (1-\lambda_2\wedge\lambda_3)^t \frac{\|X^\bot(t)\|^2}{c_x^2r_0n} + \Big(\frac{\lambda_1}{\lambda_2 \wedge \lambda_3} \wedge \lambda_1t\Big)\\\label{eq_Xbot_probbound}
        & \Big( 3 + \frac{c_l}{2} + 4 c_s c_l + \frac{2}{r_0n} \Big) \Big]. 
    \end{align}
\end{corr}
\vspace{-3ex}
\begin{rmk}
    The first result~\eqref{eq_Xgammat_probbound} bounds the probability of at least $\delta r_0n$ agents having states at least $\eps c_x$ away from the average states in their communities. This probability bound decreases first due to the decay of $(1-\lambda_3)^t$ and the relatively small value of $\lambda_i t$, and then increases with~$t$ to a constant bound (whether the constant bound is trivial depends on the ratios $\lambda_i/\lambda_3$, $i=1,2$, which we will discuss in detail in Corollary~\ref{cor_explicitlsd}). Similarly, the second result~\eqref{eq_Xbot_probbound} shows that there can be a time interval, over which many agents have states close to everyone's average state with high probability.
\end{rmk}
\vspace{-3ex}
\begin{pf}
    To prove the results, note that by the Markov inequality, for $X,Y\in \RR^{r_0n}$ and $\eps \in (0,1)$, 
    \begin{align}\nonumber
        \PP\{|\mtcs(X,Y,\eps)| \ge \delta r_0n\} &\le \PP\{\|X-Y\|^2 \ge \eps^2 \delta c_x^2 r_0n\} \\\label{eq_Sbound}
        &\le \frac{\EE\{\|X-Y\|^2\}}{\eps^2 \delta c_x^2 r_0n}.
    \end{align}
    Now note that $X(t) - X^\eta(t) - X^\xi(t) = X^\Gamma(t)$ and $X(t)-X^\eta(t) = X^\bot(t)$, so~\eqref{eq_Xgammat_probbound} and~\eqref{eq_Xbot_probbound} follow from the upper bound of $\EE\{\|X^\Gamma(t)\|^2\}$ and $\EE\{\|X^\bot(t)\|^2\}$ given in the proof of Lemma~\ref{lem_second_moment}. \hfill$\Box$
\end{pf}
\vspace{-4ex}
The preceding results indicate that we can use average states as references to describe how agent states evolve in finite time. Further analysis based on Lemma~\ref{lem_second_moment} can yield stronger results. That is, we can use the expected average states $\EE\{X^\eta(t) + X^\xi(t)\}$ and $\EE\{X^\eta(t)\}$ as references. 
{\color{black}
\begin{thm}\label{thm_deviationFromEt}
	Suppose that Assumptions~\ref{asmp_1}--\ref{asmp_3} hold and $n\ge 4/r_0$. Let $\eps,\delta,\gamma \in (0,1)$ be such that $\eps^2\delta\gamma \le 2/e$.\\
    (i) Assume that 
    \begin{align}\label{eq_lambda23log}
        &\lambda_2 \log[2/(\eps^2\delta\gamma))] < \lambda_3,\\\nonumber
        &2\Big [ (4+C_{11})\lambda_1 \log \frac{2}{\eps^2\delta\gamma}  + C_{12} \Big( 1+ \log \frac{2}{\eps^2\delta\gamma} \Big) \lambda_2 \Big] \\\label{eq_lambda123}
        &< \eps^2 \delta \gamma \lambda_3.
    \end{align}
    Then $(\underline{t}_1, \overline{t}_1) \not=\emptyset$, and for all $t\in(\underline{t}_1, \overline{t}_1)$ it holds that
    \begin{align}\label{eq_deviationEt_local}
        \PP\{|\mtcs(X(t),\EE\{X^\eta(t)+X^\xi(t)\},\eps)| \ge \delta r_0n\} \le \gamma,
    \end{align}
    where $C_{11}$ is given in Lemma~\ref{lem_second_moment}, $C_{12} = \|X^\xi(0)\|^2/(c_x^2r_0n) $ $+ 1/(r_0n)$,
    \begin{align*}
        \underline{t}_1 = \frac{\log [2/(\eps^2\delta\gamma)]}{\lambda_3},~ \overline{t}_1 = \frac{\eps^2\delta\gamma/2 - C_{12}\lambda_2/\lambda_3}{(4+C_{11})\lambda_1 + C_{12}\lambda_2} \wedge \frac{1}{\lambda_2}.
    \end{align*}
    (ii) Assume that
    \begin{align}\label{eq_lambda12wedge3}
        2\Big [(3+ C_{21}) \log \frac{2}{\eps^2\delta\gamma} + C_{22}  \Big] \lambda_1 < \eps^2 \delta \gamma (\lambda_2 \wedge \lambda_3).
    \end{align}
    Then $(\underline{t}_2, \overline{t}_2) \not=\emptyset$, and for all $t \in (\underline{t}_2, \overline{t}_2)$ it holds that
    \begin{align}\label{eq_deviationEt_global}
        \PP\{|\mtcs(X(t),\EE\{X^\eta(t)\},\eps)| \ge \delta r_0n\} \le \gamma,
    \end{align}
    where $C_{21}$ and $C_{22}$ are given in Lemma~\ref{lem_second_moment},
    \begin{align*}
        \underline{t}_2 = \frac{\log [2/(\eps^2\delta\gamma)]}{\lambda_2\wedge\lambda_3},~ \overline{t}_2 = \frac{\eps^2\delta\gamma/2 - C_{22}\lambda_1/(\lambda_2\wedge\lambda_3)}{(3+C_{21})\lambda_1}.
    \end{align*}
\end{thm}
}
\vspace{-5ex}
\begin{pf}
To prove~\eqref{eq_deviationEt_local}, from~\eqref{eq_Sbound} it suffices to bound $\EE\{\|X(t) - \EE\{X^\eta(t)\} - \EE\{X^\xi(t)\}\|^2\}$. From Lemma~\ref{lem_eigen}~(ii), we have that
\begin{align}\nonumber
    &\EE\{\|X(t) - \EE\{X^\eta(t)\} - \EE\{X^\xi(t)\}\|^2\} \\ \nonumber
    &= 
    \EE\{\|X(t)\|^2\} - 2 \EE\{X(t)\}^T (\EE\{X^\eta(t)\} + \EE\{X^\xi(t)\}) \\ \nonumber
    & + \|\EE\{X^\eta(t)\} + \EE\{X^\xi(t)\}\|^2 \\ \label{eq_pf_thmxExi_1}
    &= 
    \EE\{\|X(t)\|^2\} - \|\EE\{X^\eta(t)\} \|^2 - \|\EE\{X^\xi(t)\}\|^2.
\end{align}
{\color{black}From Lemma~\ref{lem_expression_expectation_xt} and the Bernoulli inequality, it holds that
\begin{align}\nonumber
    &\|\EE\{X^\eta(t)\}\|^2 \\ \nonumber 
    &= 
    \{(1-\lambda_1)^t \eta^T X(0) + [1-(1-\lambda_1)^t]\zeta_1\}^2\\ \nonumber
    &\ge (1-\lambda_1)^{2t} \|X^\eta(0)\|^2 - 2 \lambda_1 t (1 - \lambda_1)^t \|X^\eta(0)\| |\zeta_1| \\ \nonumber
    &\|\EE\{X^\xi(t)\}\|^2 \\ \nonumber 
    &= 
    \Big\{(1-\lambda_2)^t \xi^T X(0) + \frac{\lambda_1}{\lambda_2}[1-(1-\lambda_2)^t]\zeta_2 \Big\}^2\\ \nonumber
    &\ge (1-\lambda_2)^{2t} \|X^\xi(0)\|^2 - 2 \lambda_1 t (1 - \lambda_2)^t \|X^\xi(0)\| |\zeta_2| .
\end{align}
Thus,
\begin{align*}
    &-\|\EE\{X^\eta(t)\}\|^2-\|\EE\{X^\xi(t)\}\|^2\\
    &\le -(1-\lambda_1)^{2t} \|X^\eta(0)\|^2 - (1-\lambda_2)^{2t} \|X^\xi(0)\|^2 \\
    &+ 3\lambda_1t c_x^2r_0n
\end{align*}
Hence, when $1/\lambda_3 \le t \le 1/\lambda_2$, from~\eqref{eq_2ndmomentbound_local} in Lemma~\ref{lem_second_moment},~\eqref{eq_pf_thmxExi_1} can be bounded by
\begin{align*}
    &(1 - \lambda_3)^t \|X^\Gamma(0)\|^2 + (4+C_{11}) \lambda_1 t c_x^2 r_0n  \\
    &   + \lambda_2 \Big( \frac{1}{\lambda_3} + t \Big) [(1-\lambda_2)^t  \|X^\xi(0)\|^2 + c_x^2].
\end{align*}
Hence from~\eqref{eq_Sbound}, 
\begin{align*}
    &\PP\{|\mtcs(X(t),\EE\{X^\eta(t)+X^\xi(t)\},\eps)| > \delta r_0n\} \\
    &\le \frac{1}{\eps^2\delta} \Big[ (1-\lambda_3)^t + (4+C_{11})\lambda_1 t + C_{12} \Big( \frac{\lambda_2}{\lambda_3} + \lambda_2 t\Big) \Big].
\end{align*}
When $t \ge \underline{t}_1 \ge 1/\lambda_3$, $(1-\lambda_3)^t \le e^{t\log (1-\lambda_3)} \le e^{-\lambda_3t} \le e^{-\lambda_3\underline{t}_1} \le \eps^2\delta\gamma/2$. On the other hand, when $t \le \overline{t}_1$,
\begin{align*}
    (4+C_{11})\lambda_1 t + C_{12} \Big( \frac{\lambda_2}{\lambda_3} + \lambda_2 t\Big) \le \frac{\eps^2\delta\gamma}{2}.
\end{align*}
If the assumptions of~(i) hold, then $(\underline{t}_1,\overline{t}_1)\not=\emptyset$ and the conclusion follows. The second part of the theorem can be derived similarly from~\eqref{eq_2ndmomentbound_global}.}\hfill$\Box$
\end{pf}
\begin{rmk}
    The results provide bounds for the probability of agent states concentrating around expected average states. In particular, for the case where stubborn-agent influence is small and the influence strength within communities is much larger than that between communities ($\lambda_1$ and $\lambda_2$ much smaller than $\lambda_3$),~\eqref{eq_deviationEt_local} indicates that most agent states concentrate around expected average states within communities and form transient clusters. If the stubborn-agent influence is small but the influence strength within and between communities is similar ($\lambda_1$ much smaller than $\lambda_2\wedge\lambda_3$, and $\lambda_2$ and $\lambda_3$ are similar), then~\eqref{eq_deviationEt_global} yields that most agent states concentrate around everyone's expected average state. See Corollary~\ref{cor_explicitlsd} for detailed discussion on how link strength parameters influence the concentration.
    To obtain a nontrivial bound, $\delta$ (the proportion of regular agents whose states do not concentrate) and $\gamma$ (the probability that the concentration does not occur) need to be small. But $\eps$ (the error of the concentration) does not need. For example, it is sufficient to set $\eps$ to be less than half of the distance between average states of the two communities, so that the two transient clusters can be distinguished. For a fixed network, setting smaller $\eps$ and $\delta$ leads to a large bound of $\gamma$, as stronger concentration occurs with less probability.
    {\color{black} Since $\lambda_1$, $\lambda_2$, and $\lambda_3$ depend on the network size $n$, $n$ has to be large enough to ensure that~\eqref{eq_lambda123} and~\eqref{eq_lambda12wedge3} hold with small $\gamma$. Simulation (Figs.~\ref{fig_illus} and~\ref{fig_cases}) illustrates that such concentration can occur with high probability over small networks. Studying sharp bounds for the concentration probability would be interesting, which is left to future work.}
\end{rmk}
    
    \begin{rmk}
    In this remark we compare the obtained results with existing transient behavior analysis. The authors of~\cite{dietrich2016transient} study a generalized HK model and define a transient cluster as a subgroup of agents whose opinion range decreases faster than the distance of the subgroup from other agents. Here we characterize agent states based on their distance from the average states. Such a framework includes situations where clusters approach each other, but defining such references requires specifying subgroups beforehand. The paper~\cite{s2022finite} provides bounds for opinion difference in finite time for a stochastic bounded confidence model. These bounds coincide with the asymptotic behavior. In contrast, here we study transient behavior of the gossip model that is different from the asymptotic behavior. For the gossip model without stubborn agents,~\cite{fagnani2008randomized} shows that $\|X(t) - X^\eta(t)\|^2$ is close to its expectation at the early stage of the process, which provides insight into why both average states and expected average states can be used as references. We will study such concentration in the future.
\end{rmk}

It is possible to derive bounds similar to Theorem~\ref{thm_deviationFromEt} for $\mtcs(X(t),X^\eta(0)+X^\xi(0),\eps)$ and $\mtcs(X(t),X^\eta(0),\eps)$, i.e., the deviation of agent states from average intial states. 
We state the results in the next theorem following the notation of Theorem~\ref{thm_deviationFromEt}.

{\color{black}
\begin{thm}\label{thm_deviationInterval}
	Suppose that Assumptions~\ref{asmp_1}--\ref{asmp_3} hold and $n\ge 4/r_0$. Let $\eps,\delta,\gamma \in (0,1)$ be such that $\eps^2\delta\gamma \le 2/e$.\\
    (i) If~\eqref{eq_lambda23log} and~\eqref{eq_lambda123} hold, 
    then we have that for all $t \in (\underline{t}_1, \overline{t}_1) \not=\emptyset$, where $\underline{t}_1$ and $\overline{t}_1$ are given in Theorem~\ref{thm_deviationFromEt}~(i),
    \begin{align}\label{eq_concentrationX0_local}
        \PP\{|\mtcs(X(t),X^\eta(0)+X^\xi(0),\eps)| \ge \delta r_0n \} \le \gamma.
    \end{align}
    (ii) If~\eqref{eq_lambda12wedge3} holds, 
    then we have that for all $t \in (\underline{t}_2, \overline{t}_2) \not=\emptyset$, where $\underline{t}_2$ and $\overline{t}_2$ are given in Theorem~\ref{thm_deviationFromEt}~(ii),
    \begin{align}\label{eq_concentrationX0_global}
        \PP\{|\mtcs(X(t),X^\eta(0),\eps)| \ge \delta r_0n \} \le \gamma.
    \end{align}
\end{thm}
}

\vspace{-4ex}
\begin{pf}
The proof is similar to that of Theorem~\ref{thm_deviationFromEt}. {\color{black}See Appendix~\ref{sec_proofs_thmdevInt} for the details.} \hfill$\Box$
\end{pf}

\begin{rmk}
    {\color{black}Theorems~\ref{thm_deviationFromEt} and~\ref{thm_deviationInterval} provide similar probability bounds for concentration of agent states around expected average states and around average initial states, respectively. However, concentration around expected average states may require a much smaller network size than the latter, as illustrated in Section~\ref{sec_simulation}. Future work will explore sharp bounds to distinguish between the two types of concentration. }
\end{rmk}


So far we have studied concentration around average states based on conditions of $\lambda_i$, The following corollary, which is a  consequence of Theorem~\ref{thm_deviationInterval}, explicitly shows how such phenomena depend on link strength parameters $l_s^{(r)}$, $l_d^{(r)}$, and $l^{(s)}$. Parallel results also hold for the expected average states.

\begin{corr}\label{cor_explicitlsd}
    Suppose that Assumptions~\ref{asmp_1}--\ref{asmp_3} hold and denote $l_0^{(s)} \Let l^{(s)}/(r_0n)$.\\
    (i) If $l_s^{(r)} = \omega(l_d^{(r)} \vee l_0^{(s)})$, then for large enough~$n$ depending on $l_s^{(r)}$, $l_d^{(r)}$, and $l^{(s)}$,~\eqref{eq_concentrationX0_local} holds for all $t \in (\underline{t}_1, \overline{t}_1)$ with
    \begin{align*}
        \underline{t}_1 
        = \Big(\Theta(1) + O\Big(\frac{l_0^{(s)}}{l_s^{(r)}}\Big) \Big) r_0n,~
        \overline{t}_1 = \Theta\Big(\frac{l_s^{(r)}}{l_d^{(r)} \vee l_0^{(s)}} \Big) r_0n.
    \end{align*}
    (ii) If $l_s^{(r)} \le l_d^{(r)} + o(l_d^{(r)})$ and $l_d^{(r)} = \omega(l_0^{(s)})$, then for large enough~$n$ depending on $l_d^{(r)}$ and $l^{(s)}$,~\eqref{eq_concentrationX0_global} holds for all $t \in (\underline{t}_2, \overline{t}_2)$ with
    \begin{align*}
        \underline{t}_2 &= \Big( \Theta(1) + O\Big(\frac{[0\vee (l_s^{(r)} - l_d^{(r)})] + l_0^{(s)}}{l_d^{(r)}} \Big) \Big) r_0n,\\
        \overline{t}_2 &= \Theta\Big(\frac{l_d^{(r)}}{l_0^{(s)}} \Big) r_0n.
    \end{align*}
\end{corr}
\vspace{-6ex}
\begin{pf}
    We follow the notation of  Theorems~\ref{thm_deviationFromEt} and~\ref{thm_deviationInterval}. From definition of $\lambda_i$, it follows that~\eqref{eq_lambda23log} and~\eqref{eq_lambda123} are equivalent to
    \begin{align}\label{eq_phasetransition_lsgeld0}
        &2 \Big( \log \frac{2}{\eps^2\delta\gamma} - 1\Big) l_0^{(s)} + \Big( 2 \log \frac{2}{\eps^2\delta\gamma} - 1 \Big) l_d^{(r)} < l_s^{(r)}\\\nonumber
        &~2\Big[ 2 (4+C_{11}+C_{12}) \log \frac{2}{\eps^2\delta\gamma} + 2 C_{12} - \eps^2 \delta\gamma \Big] l_0^{(s)}\\ \label{eq_phasetransition_lsgeld}
        &+ \Big[4 C_{12} \Big(1+\log\frac{2}{\eps^2\delta\gamma} \Big) - \eps^2\delta\gamma \Big] l_d^{(r)} < \eps^2\delta\gamma l_s^{(r)}.
    \end{align}
    The condition $l_s^{(r)} = \omega(l_d^{(r)} \vee l_0^{(s)})$ guarantees that~\eqref{eq_phasetransition_lsgeld0} and~\eqref{eq_phasetransition_lsgeld} hold for large enough~$n$ that depends on $l_s^{(r)}$, $l_d^{(r)}$, and $l^{(s)}$. Hence the conclusion follows from the expression of $\underline{t}_1$ and $\overline{t}_1$ given in Theorem~\ref{thm_deviationFromEt}. The proof of~(ii) is similar. It suffices to note that~\eqref{eq_lambda12wedge3} is equivalent to
    \begin{align}\nonumber
        &~2\Big[ 2 (3+C_{21}) \log \frac{2}{\eps^2\delta\gamma} + 2 C_{22}  - \eps^2 \delta\gamma \Big] l_0^{(s)}\\ \label{eq_phasetransition_ls=ld}
        & < \eps^2\delta\gamma [l_d^{(r)} + (l_d^{(r)} \wedge l_s^{(r)})].
    \end{align}
    The condition $l_d^{(r)} = \omega(l_0^{(s)})$ ensures that~\eqref{eq_phasetransition_ls=ld} holds with large enough $n$ that depends on $l_d^{(r)}$ and $l^{(s)}$. \hfill$\Box$
\end{pf}


\begin{rmk}
	The first part of the corollary indicates that, if the intra-community weights are of higher order than the inter-community weights and the average weight between regular and stubborn agents, then most agent states concentrate around the initial average opinion of the corresponding community with high probability over a finite time interval. The length of this interval depends on relative strength within communities to that between communities and between regular and stubborn agents. The assumption $l_s^{(r)} \le l_d^{(r)} + o(l_d^{(r)})$ of the second part means that intra-community weights are less than or slightly larger than inter-community weights. If stubborn influence is also small, then most agent states are close to everyone’s initial average opinion with large probability at the early stage of the process. Note that $l_s^{(r)} \le l_d^{(r)} + o(l_d^{(r)})$ implies $l_s^{(r)} = O(l_d^{(r)})$, so the corollary characterizes a phase transition phenomenon.
\end{rmk}

The preceding results study the case where the influence of stubborn agents is small. In the following, we investigate the case with large stubborn influence.
The gossip model~\eqref{eq_update_compact_regular} converges to a unique stationary distribution $\pi$ (that is, if $X(0)$ has distribution $\pi$, then $X(t)$ has the same distribution for all $t \in \NN^+$), if there is at least one stubborn agent and the network is connected~\cite{xing2023community,acemouglu2013opinion}. 
To characterize the distance between transient and stationary distributions of agent states, we introduce the Wasserstein metric between two measures $\mu$ and $\nu$,
\begin{align*}
	d_W(\mu,\nu) \Let \inf_{(X,Y)\in J} \EE\{\|X-Y\|\},
\end{align*}
where $J$ is the set of pairs of random vectors $(X,Y)$ such that the marginal distribution of $X$ is $\mu$ and the marginal distribution of $Y$ is $\nu$. The following theorem provides the time interval over which the distribution of $X(t)$ is close to $\pi$.

\begin{table*}[ht]
\centering
\begin{center}
\caption{Summary of Main Results}\label{tb_mainrst}
\begin{tabular}{>{\centering\arraybackslash}p{0.7\columnwidth}|>{\centering\arraybackslash}p{0.9\columnwidth}|>{\centering\arraybackslash}p{0.3\columnwidth}}\hline
Conditions &  Behavior of Agent States &  Results \\\hline
$l_s^{(r)} = \omega(l_d^{(r)}\wedge l^{(s)}_0)$ with $l_0^{(s)} = l^{(s)}/(r_0n)$\newline (stubborn influence is small, \newline intra-community influence is large) & Concentrate around average states/expected average states/average initial states within communities over the time interval $(\Theta(r_0n), \Theta(l_s^{(r)}r_0n/(l_d^{(r)} \vee l_0^{(s)}))) $ & \eqref{eq_Xgammat_probbound},~\eqref{eq_deviationEt_local}, Theorem~\ref{thm_deviationInterval}~(i),\newline Corollary~\ref{cor_explicitlsd}~(i)~~~\\\hline
$l_s^{(r)} \le l_d^{(r)} + o(l_d^{(r)})$ and $l_d^{(r)} = \omega(l_0^{(s)})$\newline (stubborn influence is small, intra- \newline community influence is moderate or small) & Concentrate around everyone's average state/expected average state/average initial state over the time interval 
 $(\Theta(r_0n), \Theta(l_d^{(r)}r_0n/l_0^{(s)}))) $ & \eqref{eq_Xbot_probbound},~\eqref{eq_deviationEt_global}, Theorem~\ref{thm_deviationInterval}~(ii),\newline Corollary~\ref{cor_explicitlsd}~(ii)~~\\ \hline
$l_s^{(r)} \vee l_d^{(r)} = o(l_0^{(s)})$ \newline (stubborn influence is large) & Close to the stationary distribution over the time interval $(\Theta(r_0 n\log (r_0n)),+\infty)$& Corollary~\ref{cor_prob3}~~~~\\\hline
\end{tabular}
\end{center}
\end{table*}

\begin{thm}\label{thm_prob3}
	Suppose that Assumptions~\ref{asmp_1}--\ref{asmp_2} hold.  Then, for $\eps \in (0,1)$, it holds that 
	\begin{align*}
		d_W(X(t),\pi) \le \eps, ~\forall t > \frac{\log \{c_x (r_0 n)^{\frac52} [1 + 1/(2\lambda_1)] /\eps\}}{\log [1/(1-\lambda_1)]}.
	\end{align*}
\end{thm}

\begin{pf}
	The conclusion is obtained from a coupling argument. See Appendix~\ref{sec_proofs_thm_prob3}. \hfill$\Box$
\end{pf}


Theorem~\ref{thm_prob3} provides a general characterization for behavior of the gossip model in the long run. The following corollary focuses on the case where the influence of stubborn agents is large.

\begin{corr}\label{cor_prob3}
	Suppose that Assumptions~\ref{asmp_1}--\ref{asmp_2} hold. If $(l_d^{(r)} \vee l_s^{(r)})n = o(l^{(s)})$, then $d_W(X(t),\pi) \le \eps$ holds for $\eps \in (0,1)$ and $t > t_0$ with $t_0 = t_0(\eps) = \Theta(r_0 n\log (r_0n))$.
\end{corr}

\begin{rmk}
	Corollary~\ref{cor_prob3} indicates that, if the influence of stubborn agents is large, then the distribution of $X(t)$ can be close to the stationary distribution of the gossip model at the early stage of the process. 
\end{rmk}

\subsection{Discussion and Extension}\label{sec_dis}

In this subsection, we first summarize obtained transient behavior under different parameter settings, then discuss the extensions of the results.

In the previous subsection, we obtained several probability bounds for agent states concentrating around average states (Corollary~\ref{cor_deviationfromTimet}), expected average states (Theorem~\ref{thm_deviationFromEt}), and average initial states (Theorem~\ref{thm_deviationInterval}). Note that these bounds are the same except for some constants. So the explicit dependence of transient behavior on link strength parameters $l_s^{(r)}$, $l_d^{(r)}$, and $l^{(s)}$, given in Corollary~\ref{cor_explicitlsd}, still holds for the cases of average states and expected average states. Table~\ref{tb_mainrst} summarizes the findings. The results indicate a phase transition phenomenon: the model behaves differently at the early stage of the process under different parameter settings.

We note that the obtained bounds are not tight. As shown in Section~\ref{sec_simulation}, expected average states are good references for transients. The concentration occurs for small networks. In contrast, concentration around average initial states requires much larger $n$. 
The key idea of studying transient behavior in this paper is to find references such as the expected average states and to bound the deviation of agent states from such references. Therefore, it is possible to generalize the obtained results to multiple-community cases. With the help of matrix perturbation theory, it is also possible to deal with the case where communities do not have equal size or the total influences of stubborn agents on regular agents are different (i.e., Assumption~\ref{asmp_1}~(iii) does not hold). Studying the gossip model over an SBM needs analysis of the deviation of the random graph from its expected graph based on concentration inequalities (e.g.~\cite{chung2011spectra}).

\begin{figure*}[t]
	\centering
	\subfigure[\label{phase_transition1} The case where $l_s^{(r)}$ varies. $\beta_2 = 5$ and $\beta_3 = 1$. 
    ]{\qquad~~ \includegraphics[scale=0.4]{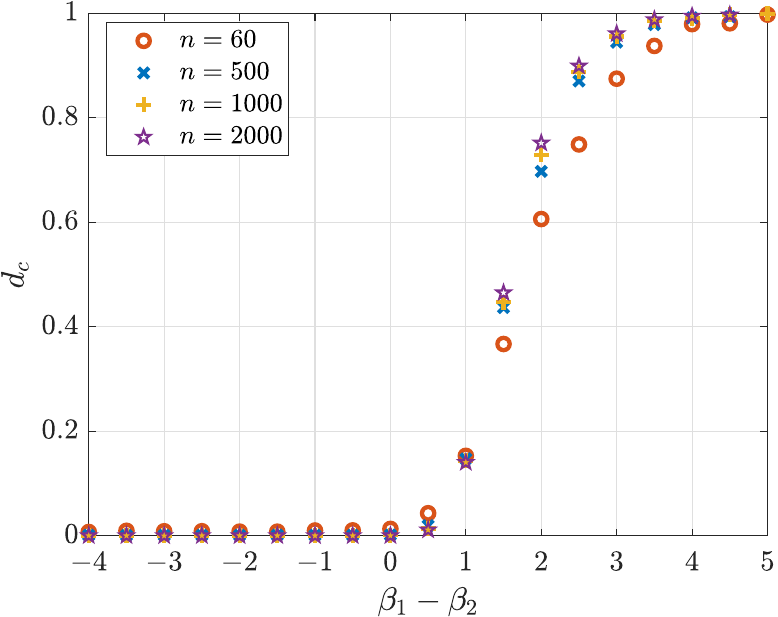}\qquad~~ } ~~
	\subfigure[\label{phase_transition2} The case where $l^{(s)}$ varies. $\beta_1 = 1$ and $\beta_2 = 5$. 
    ]{\qquad~~\includegraphics[scale=0.4]{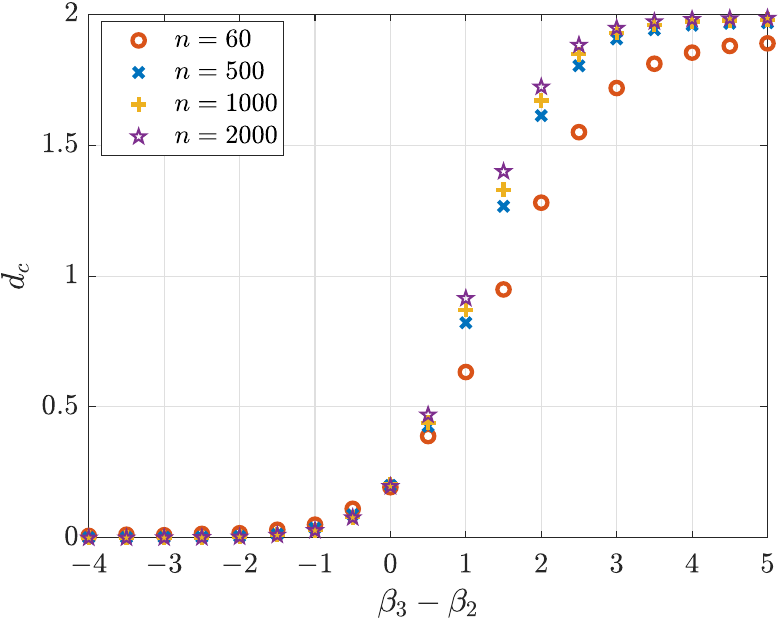}\qquad~~}
	\caption{\label{phase_transition} Phase transition of transient behavior of the gossip model. In the experiments, we set $l_s^{(r)} = (\log^{\beta_1} n)/n$, $l_d^{(r)} = (\log^{\beta_2} n)/n$, and $l^{(s)} = \log^{\beta_3} n$, and compute $d_c = 2 |\sum_{i \in \mtcv_{r1}} X_i(t) - \sum_{i \in \mtcv_{r2}} X_i(t)| / (r_0n)$ with $t = \lfloor n\log n\rfloor$.}	
\end{figure*}

\section{Numerical Simulation}\label{sec_simulation}
In this section, we conduct numerical simulations to validate the theoretical results obtained in Section~\ref{sec_results}. 



In the first set of experiments shown in Fig.~\ref{fig_illus}, we demonstrate transient behavior of the gossip model. 
Set the size of the network $n=60$, the bound of agent states $c_x = 1$, and the proportion of regular agents $r_0 = 0.9$. Generate the initial value $X_i(0)$ independently from uniform distribution on $(0,1)$ for all $i \in \mtcv_{r1}$, $X_j(0)$ independently from uniform distribution on $(-1,0)$ for all $j \in \mtcv_{r2}$, and set stubborn-agent states $z^s = [\bfl_{s_0n/2}~ -\bfl_{s_0n/2}]$. 
First, we set the edge weight between agents in the same community to be $l_s^{(r)} = (\log^{2.5} n)/n$, greater than the edge weight between communities $l_d^{(r)} = (\log  n)/n$. The edge weights between regular and stubborn agents are as follows: for all $i\in \mtcv_{r1}$, $l_{ij}^{(s)} \equiv (\log n)/n$ for $1\le j \le s_0n/2$, and $a_{ij} \equiv 0$ for $1+s_0n/2\le j \le s_0n$; for all $i\in \mtcv_{r2}$, $l_{ij}^{(s)} \equiv 0$ for $1\le j \le s_0n/2$, and $a_{ij} \equiv (\log n)/n$ for $1+s_0n/2\le j \le s_0n$. Hence $l^{(s)} = (s_0 \log n)/2$. This setting intuitively means that the first half of stubborn agents influence only $\mtcv_{r1}$, whereas the second half influence only $\mtcv_{r2}$. We use this setting for all numerical experiments. Fig.~\ref{fig_illus_case1a} shows that agents form two transient clusters corresponding to their community labels and centered around the expected average states within communities.
Asymptotic behavior of the system can be different from what shown in Fig.~\ref{fig_illus_case1a}. As an example, set  $l_s^{(r)} = (\log^{2.5} n)/n$, $l_d^{(r)} = (\log  n)/n$, and $l^{(s)} = (s_0 \log^2 n)/2$. As in Fig.~\ref{fig_illus_case1b}, agents in the same community also form transient clusters, but they get close in the end. 
Next, we set $l_s^{(r)} = (\log^{2.5} n)/n$, $l_d^{(r)} = (\log^{2.4}  n)/n$, and $l^{(s)} = (s_0 \log n)/2$.  In this case, the influence of stubborn agents is small, and inter-community influence strength is similar to intra-community strength. Hence all agents form a single transient cluster and concentrate around everyone's average state, as shown in Fig.~\ref{fig_illus_case2}. However, they cannot reach a  consensus and keep fluctuating in the long run, because of the presence of stubborn agents.
Finally, set $l_s^{(r)} = (\log n)/n$, $l_d^{(r)} = (\log n)/n$, and $l^{(s)} = (s_0 \log^3 n)/2$. Fig.~\ref{fig_illus_case3} illustrates the case where the influence of stubborn agents is large. The figure shows that agent states move quickly towards the positions of stubborn agents. 

Next we study the probability of concentration around average states, investigated in Theorems~\ref{thm_deviationFromEt} and~\ref{thm_deviationInterval}. We first study concentration around expected average states. Set $n=60$, $l_s^{(r)} = (\log^{2.5} n)/n$, $l_d^{(r)} = (\log  n)/n$, and $l^{(s)} = (s_0 \log n)/2$ (i.e., intra-community edge weights are large). The model is run for $200$ times and the final time step is $\lfloor n(\log n)^2\rfloor$. We estimate the probability $p_t$ that the concentration fails by computing $\frac{1}{200}\sum_{k=1}^{200} \mathbb{I}_{[|\mtcs_t^\eps(k)| \ge \delta r_0 n]}$. Here $\mtcs_t^\eps(k)$ is the set $\mtcs(X(t),\EE\{X^\eta(t)+X^\xi(t)\},\eps)$ at the $k$-th run, that is, the set of agents with states at least $\eps c_x$ away from the expected average state of their communities at time~$t$. The concentration error $\eps$ and the proportion of non-concentrating agents $\delta$ are defined in Theorem~\ref{thm_deviationFromEt}.
Fig.~\ref{fig_expcase1} shows that $p_t$ first decreases with time and then fluctuates around a constant, and it decreases with $\delta$. These findings validate the bound~\eqref{eq_deviationEt_local}.
Although the bound~\eqref{eq_deviationEt_local} is not tight (e.g., the ratio $\lambda_2/(\lambda_3 \eps^2 \delta) \approx 18.7>1$), the experiment shows that the concentration still occurs for small networks. Next, we set $l_d^{(r)} = (\log^{2.4}  n)/n$ for the case of moderate intra-community influence. Fig.~\ref{fig_expcase2} illustrates that the probability varies similarly, validating the bound~\eqref{eq_deviationEt_global}. Now we examine the concentration around average initial states. Fig.~\ref{fig_initcase1_60} shows that the concentration around average initial states within communities happens with less probability than that around expected average states shown in Fig.~\ref{fig_expcase1}. Larger networks ensure more concentration, as shown in Fig.~\ref{fig_initcase1_500} with $n=500$. The concentration around everyone's average initial state, given in Fig.~\ref{fig_initcase2}, is similar to the expected state case.

Finally, for concentration around average initial states, we conduct two experiments to demonstrate how the phase transition occurs when edge weights vary. In the first simulation, we set $l_d^{(r)} = (\log^{\beta_2} n)/n = (\log^5 n)/n$ and $l^{(s)} = \log n$, and consider $l_s^{(r)} = (\log^{\beta_1} n)/n$ with $\beta_1 = 1,\dots, 10$. We run the gossip model for $50$ times, for each value of $\beta_1$ and for $n=60,500,1000, 2000$, and compute the difference $d_c = 2 |\sum_{i \in \mtcv_{r1}} X_i(t) - \sum_{i \in \mtcv_{r2}} X_i(t)| / (r_0n)$ with $t = \lfloor n \log n\rfloor$. 
The difference $d_c$ represents the difference between the averages of agent states in the two communities. The case $d_c = 1$ means local concentration whereas $d_c = 0$ represents global concentration. Fig.~\ref{phase_transition1} presents the phase transition phenomenon (the effect becomes stronger as $n$ grows). It shows that the local concentration appears when $\beta_1-\beta_2 > 0$ (i.e., $l_d^{(r)}  =o( l_s^{(r)})$), as predicted by Corollary~\ref{cor_explicitlsd}. On the other hand, Fig.~\ref{phase_transition1} indicates that the global concentration appears when $\beta_1-\beta_2 < 0$. In the second simulation, we set $l_s^{(r)} = (\log  n)/n$ and $l_d^{(r)} = (\log^{\beta_2} n)/n = (\log^5 n)/n$, and consider $l^{(s)} = \log^{\beta_3} n$ with $\beta_3 = 1,\dots, 10$. This simulation examines the phase transition in the influence of stubborn agents. Under this circumstance, $d_c = 2$ means regular agents have states close to stubborn ones, whereas $d_c = 0$ represents the global concentration. Fig.~\ref{phase_transition2} illustrates that the transition occurs at $n l_d^{(s)} = \Theta(l^{(s)})$ (i.e., $\beta_3 = \beta_2$), as predicted by Corollary~\ref{cor_prob3}.

\section{Conclusion}\label{sec_conclusion}
In this paper, we investigated transient behavior of the gossip model with two communities. By analyzing the second moment of agent states, we found a phase transition phenomenon: When edge weights within communities are large and weights between regular and stubborn agents are small, most agents in the same community have states close to the average opinion of that community at the early stage of the process. When the difference between intra- and inter-community weights is small or the inter-community weights are larger than the intra-community ones, most agents in the network have states close to everyone's average opinion. In contrast, if weights between regular and stubborn agents are large, the distribution of agent states is close to the stationary distribution. Future work includes to study the gossip model over general graphs and other models, and to link theoretical findings to empirical data.

\section*{Acknowledgment}
This work was supported by the Knut \& Alice Wallenberg Foundation and the Swedish Research Council. The authors would like to thank the anonymous reviewers for their insightful comments and suggestions.

\textbf{Appendix}
\appendix
\section{Proof of Lemma~\ref{lem_second_moment}}\label{sec_append_proof_lem3}

\textbf{PROOF OF~\eqref{eq_2ndmomentbound_local}.}  
We divide the proof of~\eqref{eq_2ndmomentbound_local} into four steps. Step~1 provides an upper and a lower bound for $\EE\{\|X^{\eta}(t)\|^2\}$. Bounds of $\EE\{\|X^{\xi}(t)\|^2\}$ are given in step~2. The lower bounds of $\EE\{\|X^{\eta}(t)\|^2\}$ and $\EE\{\|X^{\xi}(t)\|^2\}$ are used in step~3, obtaining an upper bound for $\EE\{\|X^{\Gamma}(t)\|^2\}$. Step~4 puts everything together and completes the proof.
	
\noindent\textbf{Step 1 (Upper and lower bounds of $\EE\{\|X^{\eta}(t)\|^2\}$).} Note that
\begin{align}\nonumber
	&\EE\{\| X^{\eta}(t+1)\|^2 | X(t)\}\\\nonumber 
	&=  
	\EE\{ X(t+1)^T \eta\eta^T \eta\eta^T X(t+1) | X(t)\} \\\nonumber
	&= 
	\EE\{X(t)^T Q(t)^T \eta \eta^T Q(t) X(t) | X(t)\} + 2\EE\{  X(t)^TQ(t)^T \\\nonumber
	&~ \eta \eta^T R(t)z^s | X(t)\} + \EE\{  (z^s)^TR(t)^T \eta \eta^T R(t)z^s | X(t)\} \\\nonumber 
	&=
	X(t)^T \EE\{ Q(t)^T \eta \eta^T Q(t)\}  X(t)  + 2X(t)^T \EE\{  Q(t)^T \eta \eta^T \\\label{eq_complete_secondmoment_x_eta1}
	&R(t) \} z^s + (z^s)^T\EE\{  R(t)^T \eta \eta^T R(t) \}z^s.
\end{align}
In the last equation, $X(t)$ and $z^s$ are taken outside the conditional expectations because of measurability, and the conditional expectations degenerate into expectations because $[Q(t),R(t)]$ is independent of $X(t)$. Let $\mtce_r$ be the collection of edges whose end points are regular agents, and let $\mtce_s$ be the set of edges connecting regular and stubborn agents. Hence, $\mtce = \mtce_r \cup \mtce_s$. From the definition of $Q(t)$, we have that, at time $t$, $Q(t)^T\eta \eta^T Q(t) = \eta \eta^T$ when an edge in $\mtce_r$ is selected, and $Q(t)^T\eta \eta^T Q(t)=  (\eta - \eta_u \bfe_u /2)(\eta - \eta_u \bfe_u /2)^T$ when $\{u,v\} \in \mtce_s$ is selected. Put it together, we get that 
\begin{align*}
	&\EE\{ Q(t)\eta \eta^T Q(t) \} \\
	&=
	\eta \eta^T +  \frac{l^{(s)}}{2\alpha} \sum_{u \in \mtcv_r} (-\eta_u \bfe_u)\eta^T + \frac{l^{(s)}}{2\alpha} \sum_{u \in \mtcv_r} \eta (-\eta_u \bfe_u)^T \\\nonumber
        & + \frac{l^{(s)}}{4\alpha} \sum_{u \in \mtcv_r}  \eta_u^2 \bfe_u \bfe_u^T \\
        &=
	(1 - 2\lambda_1) \eta \eta^T + \frac{\lambda_1}{2r_0n} I_{r_0n},
\end{align*}
where $e_i \Let e^{(r_0n)}_i$. Denote $e^{s}_i \Let e^{(s_0n)}_i$, $\tilde{l}^{(s)}_v \Let \sum_{u\in \mtcv_r} l_{uv}^{(s)}$, and $\tilde{\eta} \Let \sum_{1\le v\le s_0n} \tilde{l}_v^{(s)} e_v^s/\sqrt{s_0n}$. Then
\begin{align*}
	&\EE\{  Q(t)^T \eta \eta^T R(t) \} \\
        &= 
	\sum_{\{u,v+r_0n\} \in \mtce_s} \Big(\eta - \frac12\eta_u \bfe_u\Big) \Big(\frac12 \eta_u e^{s}_v \Big)^T \frac{l_{uv}^{(s)}}{\alpha}\\
	&= 
	\sum_{1\le v\le s_0n} \sum_{u \in \mtcv_r} \Big( \frac{l_{uv}^{(s)}}{2\alpha} \eta_u \eta (e_v^s)^T -   \frac{l_{uv}^{(s)}}{4\alpha} \eta_u^2 \bfe_u (e_v^s)^T \Big) \\
	&=
	\frac{c_s \lambda_1 }{l^{(s)}} \eta \tilde{\eta}^T - \frac{\lambda_1}{2 r_0n l^{(s)}}   \tilde{M},  \\
	&\EE\{ R(t)^T \eta \eta^T R(t) \} \\
	&=
	\frac14 \sum_{1\le v\le s_0n} \sum_{u \in \mtcv_r} \frac{l_{uv}^{(s)}}{\alpha} \eta_u^2 e^s_v (e^s_v)^T 
        =
	\frac{\lambda_1 \tilde{D}^{(s)}}{2 r_0 n l^{(s)}} ,
\end{align*}
where $\tilde{D}^{(s)} \Let \sum_{1\le v\le s_0n} \tilde{l}_{v}^{(s)} e^s_v (e^s_v)^T$.
Hence,~\eqref{eq_complete_secondmoment_x_eta1} is 
\begin{align}\nonumber
	& 
	X(t)^T \Big[ (1 - 2\lambda_1) \eta \eta^T + \frac{\lambda_1}{2r_0n} I_{r_0n} \Big] X(t) + 2 X(t)^T\\\nonumber
        & \Big[  \frac{c_s\lambda_1 }{l^{(s)}} \eta \tilde{\eta}^T - \frac{\lambda_1}{2 r_0n l^{(s)}} \tilde{M} \Big] z^s + (z^s)^T \frac{\lambda_1 \tilde{D}^{(s)}}{2 r_0n l^{(s)}}  z^s\\\nonumber
	&= 
	(1 - 2 \lambda_1) \|X^{\eta}(t)\|^2 + \frac{\lambda_1}{2r_0n} \|X(t)\|^2 + \frac{2c_s \lambda_1}{l^{(s)}}  X(t)^T \\\nonumber
        & \eta \tilde{\eta}^T z^s  -  \frac{\lambda_1}{ r_0n l^{(s)}} X(t)^T  \tilde{M} z^s +  \frac{\lambda_1 }{2 r_0n l^{(s)}} (z^s)^T \tilde{D}^{(s)} z^s.
\end{align}
Taking expectation yields the following upper and lower bounds for $\EE\{\|X^{\eta}(t)\|^2\}$ when $n \ge 1/r_0$,
\begin{align}\nonumber
	&\EE\{\|X^{\eta}(t+1)\|^2\} \\\nonumber
        &\le \Big(1 - \frac32 \lambda_1\Big) \EE\{\|X^{\eta}(t)\|^2\} + \frac{\lambda_1}{2r_0n} (\EE\{\|X^{\xi}(t)\|^2\} \\\nonumber
        & + \EE\{\|X^{\Gamma}(t)\|^2\})  + \frac{2c_s\lambda_1 }{l^{(s)}}   \EE\{X(t)\}^T \eta \tilde{\eta}^T z^s \\\label{eq_complete_secondmoment_x_eta_upperbound_step1}
	& -  \frac{\lambda_1}{ r_0n l^{(s)}} \EE\{X(t)\}^T  \tilde{M} z^s + \frac{\lambda_1 \tilde{l}^{(s)}_+}{2 r_0 n l^{(s)}} \|z^s\|^2,\\\nonumber
	&\EE\{\|X^{\eta}(t+1)\|^2\} \ge (1 - 2 \lambda_1) \EE\{\|X^{\eta}(t)\|^2\}+ \frac{2c_s\lambda_1}{l^{(s)}}\\\label{eq_complete_secondmoment_x_eta_lowerbound_step1}
        &      \EE\{X(t)\}^T \eta\tilde{\eta}^T z^s  -  \frac{\lambda_1}{ r_0n l^{(s)}} \EE\{X(t)\}^T  \tilde{M} z^s.
\end{align}
Furthermore, by induction and from~\eqref{eq_complete_secondmoment_x_eta_upperbound_step1}, Assumptions~\ref{asmp_2} and~\ref{asmp_3}, and Lemma~\ref{lem_append_extrabounds} at the end of this section,
\begin{align}\nonumber
    &\EE\{\|X^{\eta}(t)\|^2\}
    \le
    (1-\lambda_1)^t \|X^\eta(0)\|^2 + \lambda_1 \Big(\frac{1}{\lambda_1} \wedge t\Big) \\\nonumber& \Big[ 2c_s c_l  (\|X^\eta(0)\| + |\zeta_1|) \|z^s\| + \frac{1}{2r_0n} (c_x^2r_0n + \|X(0)\|^2 \\ \label{eq_complete_secondmoment_x_eta_upperbound2_step1}
    & + 3 \|\tilde{z}^s\|^2 + c_l \|z^s\|^2) \Big],
\end{align}
where $\tilde{z}^s \Let \tilde{M}z^s/l^{(s)}$ and $\zeta_1 \Let \eta^T \bar{R} z^s/\lambda_1 =  \eta^T \tilde{M} z^s / l^{(s)}$.

\noindent\textbf{Step 2 (Upper and lower bounds of $\EE\{\|X^{\xi}(t)\|^2\}$).} 
For $X^{\xi}(t)$, it holds that
\begin{align}\nonumber
	&\EE\{\|X^{\xi}(t+1)\|^2 | X(t)\} \\\nonumber 
	&= 
	\EE\{X(t)^T Q(t)^T \xi \xi^T Q(t) X(t) | X(t)\} + 2\EE\{  X(t)^TQ(t)^T \\\nonumber
        & \xi \xi^T R(t)z^s | X(t)\}  + \EE\{  (z^s)^TR(t)^T \xi \xi^T R(t)z^s | X(t)\} \\ \nonumber
	&=
	X(t)^T \EE\{ Q(t)^T \xi \xi^T Q(t)\}  X(t) + 2X(t)^T \EE\{  Q(t)^T \xi \xi^T \\\label{eq_complete_secondmoment_x_xi1}
        & R(t) \} z^s  + (z^s)^T\EE\{  R(t)^T \xi \xi^T R(t) \}z^s,
\end{align}
where the last equation is obtained in the same way as~\eqref{eq_complete_secondmoment_x_eta1}. For $Q(t)^T \xi\xi^T$$Q(t)$, we have that
\begin{align*}
	&\EE\{Q(t)^T\xi\xi^T Q(t)\} \\
	&=
	\underset{\mtcc_u = \mtcc_v}{\sum_{\{u,v\} \in \mtce_r}} \frac{l_s^{(r)}}{\alpha} \xi \xi^T + \underset{\mtcc_u \not= \mtcc_v}{\sum_{\{u,v\} \in \mtce_r}} \frac{l_d^{(r)}}{\alpha} (\xi - \xi_u e_u - \xi_v e_v) \\\nonumber
	& (\xi - \xi_u e_u - \xi_v e_v)^T + \sum_{\{u,v\} \in \mtce_s} \frac{l_{u,v-r_0n}^{(s)}}{\alpha} \Big(\xi - \frac12 \xi_u e_u\Big) \\\nonumber
	& \Big(\xi - \frac12 \xi_u e_u\Big)^T \\
	&=
	\Big(\underset{\mtcc_u = \mtcc_v}{\sum_{\{u,v\} \in \mtce_r}} + \underset{\mtcc_u \not= \mtcc_v}{\sum_{\{u,v\} \in \mtce_r}} + \sum_{\{u,v\} \in \mtce_s} \Big) \Big( \frac{a_{uv}}{\alpha}	\xi \xi^T \Big) + \frac{l_d^{(r)}}{\alpha} \\\nonumber
	& \underset{\mtcc_u \not= \mtcc_v}{\sum_{\{u,v\} \in \mtce_r}} [-\xi (\xi_u e_u + \xi_v e_v)^T - (\xi_u e_u + \xi_v e_v)\xi^T]  + \frac{l_d^{(r)}}{\alpha}\\
	& \underset{\mtcc_u \not= \mtcc_v}{\sum_{\{u,v\} \in \mtce_r}} (\xi_u e_u + \xi_v e_v)(\xi_u e_u + \xi_v e_v)^T + \sum_{u\in \mtcv_r} \sum_{1\le v \le s_0n} \\\nonumber
	&\frac{l_{uv}^{(s)}}{2\alpha} (- \xi_u\xi e_u^T - \xi_u e_u \xi^T) + \sum_{u\in\mtcv_r} \sum_{1\le v \le s_0n} \frac{l_{uv}^{(s)}}{4\alpha} \xi_u^2 e_u e_u^T \\
	&=
	\xi\xi^T - \frac{l_d^{(r)}}{\alpha} r_0n \xi\xi^T + \frac{l_d^{(r)}}{\alpha} \underset{\mtcc_u \not= \mtcc_v}{\sum_{\{u,v\} \in \mtce_r}} (\xi_u e_u + \xi_v e_v) \\\nonumber
	& (\xi_u e_u + \xi_v e_v)^T - \frac{l^{(s)}}{\alpha} \xi\xi^T + \frac{l^{(s)}}{4\alpha r_0n} I_{r_0n} \\
	&=
	\Big( 1 - \frac{l_d^{(r)} r_0n + l^{(s)}}{\alpha} \Big)\xi\xi^T  + \frac{\lambda_1}{2 r_0 n} I_{r_0n} + \frac{2(\lambda_2-\lambda_1)}{r_0n} \\\nonumber
	& \underset{\mtcc_u \not= \mtcc_v}{\sum_{\{u,v\} \in \mtce_r}} (\xi_u e_u + \xi_v e_v)(\xi_u e_u + \xi_v e_v)^T\\
	&=
	( 1 - 2\lambda_2 )\xi\xi^T  + \frac{\lambda_1}{2 r_0 n} I_{r_0n} + \frac{2(\lambda_2-\lambda_1)}{r_0n} \Big(\xi \xi^T + \frac12 \Gamma \Big),
\end{align*}
where the last equation is obtained from
\begin{align*}
	&\underset{\mtcc_u \not= \mtcc_v}{\sum_{\{u,v\} \in \mtce_r}} (\xi_u e_u + \xi_v e_v)(\xi_u e_u + \xi_v e_v)^T\\ &= \frac12 I_{r_0n} + \sum_{u \in \mtcv_{r1}} \sum_{v\in \mtcv_{r2}} (\xi_u \xi_v e_u e_v^T + \xi_u \xi_v e_v e_u^T) \\\nonumber
	&= \frac12 I_{r_0n} + \frac12(\xi\xi^T - \eta\eta^T) = \xi \xi^T + \frac12 \Gamma.
\end{align*}
Recall that $e^s_i = e^{(s_0n)}_i$. Similary to Step~1, we have that 
\begin{align*}
	\EE\{Q(t)^T \xi \xi^T R(t)\}
	=
	\frac{c_s\lambda_1}{l^{(s)}}  \xi \tilde{\xi}^T - \frac{\lambda_1}{2 r_0n l^{(s)}}   \tilde{M}, 
\end{align*}
where we denote $\tilde{\xi} \Let \frac{1}{\sqrt{s_0n}} \sum_{1\le v\le s_0n} (\sum_{u \in \mtcv_{r1}} l_{uv}^{(s)} - \sum_{u \in \mtcv_{r2}} l_{uv}^{(s)} )e^s_v$. In addition, $\EE\{ R(t)^T \xi \xi^T R(t) \} = \lambda_1 \tilde{D}^{(s)}/(2 r_0nl^{(s)})$.
Therefore,
\begin{align*}
	&\eqref{eq_complete_secondmoment_x_xi1} = 
	  (1 - 2 \lambda_2) \|X^{\xi}(t)\|^2 +  \frac{\lambda_1}{2 r_0n} \|X^{\eta}(t)\|^2+ \frac{4\lambda_2 - 3\lambda_1}{2r_0n} \\\nonumber
        &  \|X^{\xi}(t)\|^2 + \frac{2\lambda_2 - \lambda_1}{2r_0n} \|X^{\Gamma}(t)\|^2 +  \frac{2c_s\lambda_1}{l^{(s)}}  X(t)^T \xi \tilde{\xi}^T z^s \\\nonumber
	& - \frac{\lambda_1}{r_0n l^{(s)}} X(t)^T  \tilde{M} z^s +  \frac{\lambda_1 }{2 r_0nl^{(s)}} (z^s)^T \tilde{D}^{(s)} z^s.
\end{align*}
Hence when $n \ge 4/r_0$, 
the following bounds hold
\begin{align}\nonumber
	&\EE\{\|X^{\xi}(t+1)\|^2\} \\\nonumber
        &\le
	  \Big(1 - \frac32 \lambda_2 \Big) \EE\{\|X^{\xi}(t)\|^2\} +  \frac{\lambda_1}{2 r_0n} \EE\{\|X^{\eta}(t)\|^2\} \\\nonumber
        & + \frac{\lambda_2}{r_0n} \EE\{\|X^{\Gamma}(t)\|^2\} +  \frac{2c_s\lambda_1}{l^{(s)}}  \EE\{X(t)\}^T \xi \tilde{\xi}^T z^s \\\label{eq_complete_secondmoment_x_xi_upper_bound_step2}
	&   - \frac{\lambda_1}{r_0n l^{(s)}} \EE\{X(t)\}^T  \tilde{M} z^s +  \frac{\lambda_1 \tilde{l}^{(s)}_+}{2 r_0nl^{(s)}} \|z^s\|^2,\\\nonumber 
	&\EE\{\|X^{\xi}(t+1)\|^2\} \ge
	  (1 - 2\lambda_2 ) \EE\{\|X^{\xi}(t)\|^2\}   +  \frac{2c_s\lambda_1}{l^{(s)}}\\\label{eq_complete_secondmoment_x_xi_lower_bound_step2}
        &  \EE\{X(t)\}^T \xi \tilde{\xi}^T z^s  - \frac{\lambda_1}{r_0n l^{(s)}} \EE\{X(t)\}^T  \tilde{M} z^s.
\end{align}
Furthermore, by induction and from~\eqref{eq_complete_secondmoment_x_xi_upper_bound_step2}, Assumptions~\ref{asmp_2} and~\ref{asmp_3}, and Lemma~\ref{lem_append_extrabounds} at the end of this section,
\begin{align}\nonumber
    &\EE\{\|X^\xi(t)\|^2\}
    \le (1-\lambda_2)^t \|X^\xi(0)\|^2 + \lambda_1 \Big(\frac{1}{\lambda_2} \wedge t\Big) \\\nonumber& \Big[ 2c_sc_l (\|X^\xi(0)\|  + |\zeta_2|)\|z^s\|+ \frac{1}{2r_0n} (c_x^2r_0n + \|X(0)\|^2 \\  
     \label{eq_complete_secondmoment_x_xi_upper_bound2_step2}
    & + 3\|\tilde{z}^s\|^2 + c_l\|z^s\|^2) \Big] + \lambda_2 \Big(\frac{1}{\lambda_2} \wedge t\Big) c_x^2,
\end{align}
where $\zeta_2 \Let \xi^T \bar{R} \bfz^s/\lambda_1 =  \xi^T \tilde{M} \bfz^s / l^{(s)}$.

\noindent\textbf{Step 3 (Upper bound of $\EE\{\|X^{\Gamma}(t)\|^2\}$).} 
Expanding $X(t+1)$ yields that
\begin{align}\nonumber
	&\EE\{ \|X(t+1)\|^2 |X(t)\} \\\nonumber
	&= 
	X(t)^T \EE\{Q(t)^T Q(t)\} X(t) + 2 X(t)^T \EE\{Q(t)^T R(t)\} z^s\\\label{eq_complete_secondmoment_x_gamma1}
        & + z^s \EE\{R(t)^T R(t)\} z^s.
\end{align}
Note that 
\begin{align*}
	&\EE\{Q(t)^T Q(t)\} \\
        &= 
	\sum_{\{u,v\} \in \mtce_r} \Big[ I - \frac12(\bfe_u - \bfe_v)(\bfe_u - \bfe_v)^T \Big] \frac{a_{uv}}{\alpha} \\\nonumber
        &+ \sum_{\{u,v\} \in \mtce_s} \Big( I - \frac34 \bfe_u \bfe_u^T \Big) \frac{a_{uv}}{\alpha} \\
	&=
	\bar{Q} - \frac{\lambda_1}{2} I_{r_0n}\\
	&=
	\Big(1 - \frac32 \lambda_1\Big) \eta \eta^T +  \Big( 1 - \lambda_2 - \frac12 \lambda_1\Big) \xi \xi^T \\\nonumber
        & + \Big( 1 - \lambda_3 - \frac12 \lambda_1\Big) \Gamma,\\
	&\EE\{Q(t)^TR(t)\}  \\
	&=
	\sum_{1\le v \le s_0n} \sum_{u \in \mtcv_r} \Big( I - \frac12 \bfe_u \bfe_u^T \Big) \Big(\frac12 \bfe_u (e^s_v)^T \Big) \frac{l_{uv}^{(s)}}{\alpha}
        = \frac{\lambda_1\tilde{M}}{2l^{(s)}} ,\\
	&\EE\{R(t)^T R(t)\}  \\
        &  =
        \frac{1}{4\alpha}  \sum_{1\le v \le s_0n} \sum_{u \in \mtcv_r}  e^s_v  (e^s_v)^T l_{uv}^{(s)} = 
	\frac{\lambda_1\tilde{D}^{(s)}}{2 l^{(s)}}.  
\end{align*}
Thus,
\begin{align}\nonumber
        &\eqref{eq_complete_secondmoment_x_gamma1} = \Big(1 - \frac32 \lambda_1\Big) \|X^{\eta}(t)\|^2 + \Big( 1 - \lambda_2 - \frac12 \lambda_1\Big)   \\\nonumber
        &\|X^{\xi}(t)\|^2 + \Big( 1 - \lambda_3 - \frac12 \lambda_1\Big) \|X^{\Gamma}(t)\|^2 + \frac{\lambda_1 }{l^{(s)}} X(t)^T \tilde{M} z^s \\\label{eq_usedinII}
        & + \frac{\lambda_1 }{2 l^{(s)}} (z^s)^T \tilde{D}^{(s)} z^s.
\end{align}
It follows from the preceding equation, Lemma~\ref{lem_eigen}~(ii),~\eqref{eq_complete_secondmoment_x_eta_lowerbound_step1} and~\eqref{eq_complete_secondmoment_x_xi_lower_bound_step2} that
\begin{align}\nonumber
	&\EE\{\|X^{\Gamma}(t+1) \|^2 \} \\\nonumber
	&\le 
        \lambda_1   \EE\{\|X^{\eta}(t)\|^2\} + \lambda_2   \EE\{\|X^{\xi}(t)\|^2\} + ( 1 - \lambda_3 ) \\\nonumber
        &~ \EE\{\|X^{\Gamma}(t)\|^2\} + \Big(1 + \frac{2}{r_0n} \Big) \frac{\lambda_1 }{l^{(s)}}  \EE\{X(t)\}^T \tilde{M} z^s+ \frac{\lambda_1\tilde{l}^{(s)}_+}{2 l^{(s)}}   \\\nonumber
        & \|z^s\|^2  - \frac{2c_s\lambda_1}{l^{(s)}} (  \EE\{X(t)\}^T \eta \tilde{\eta}^T z^s + \EE\{X(t)\}^T \xi \tilde{\xi}^T z^s ).
\end{align}
From Lemma~\ref{lem_append_extrabounds}, Assumptions~\ref{asmp_2} and~\ref{asmp_3}, and~\eqref{eq_complete_secondmoment_x_xi_upper_bound2_step2} and by induction, we know that
\begin{align}\nonumber
    &\EE\{\|X^\Gamma(t)\|^2\} \\ \nonumber
    &\le 
    \lambda_2 \EE\{\|X^\xi(t-1)\|^2\} + (1 - \lambda_3) \EE\{\|X^\Gamma(t-1)\|^2\} \\ \nonumber
    &  + \lambda_1 c_x^2 r_0n +  \Big(1 + \frac{2}{r_0n} \Big) \frac{\lambda_1 }{2}  (\|X(0)\|^2 + 3\|\tilde{z}^s\|^2) + \frac{c_l\lambda_1}{2}  \\ \nonumber
    &  \|z^s\|^2  + 2c_sc_l\lambda_1 \|z^s\| ( \|X^\eta(0)\| + \|X^\xi(0)\| + |\zeta_1| + |\zeta_2| ) \\  \nonumber
    &\le 
    (1-\lambda_3)^t \|X^\Gamma(0)\|^2 + \lambda_1 \Big(\frac{1}{\lambda_3} \wedge t \Big) \Big[ \Big(1+\frac{1}{2r_0n} \Big) c_x^2 r_0n \\ \nonumber
    &+ \frac{1 }{2}\Big(1 + \frac{3}{r_0n} \Big)   (\|X(0)\|^2 + 3\|\tilde{z}^s\|^2) + \Big(1+\frac{1}{r_0n} \Big) \frac{c_l}{2} \\ \nonumber
    &~ \|z^s\|^2  + 2c_sc_l \|z^s\| ( \|X^\eta(0)\| + 2\|X^\xi(0)\| + |\zeta_1| + 2|\zeta_2| ) \Big]\\ \label{eq_xgamma_upperbound2}
    & + \lambda_2 \Big(\frac{1}{\lambda_3} \wedge t \Big) [(1-\lambda_2)^t\|X^\xi(0)\|^2 + c_x^2 ].
\end{align}
\noindent\textbf{Step 4 (Putting everything together).} 
Lemma~\ref{lem_eigen} yields that $\EE\{\|X(t)\|^2\}
    = \EE\{\|X^\eta(t)\|^2\} + \EE\{\|X^\xi(t)\|^2\} + \EE\{\|X^\Gamma(t)\|^2\}$,
so~\eqref{eq_2ndmomentbound_local} follows from summarizing~\eqref{eq_complete_secondmoment_x_eta_upperbound2_step1},~\eqref{eq_complete_secondmoment_x_xi_upper_bound2_step2}, and~\eqref{eq_xgamma_upperbound2} and using the fact $1/\lambda_1 \ge 1/\lambda_k$, $k=2,3$.

{\color{black}\textbf{PROOF OF~\eqref{eq_2ndmomentbound_global}.} 
The derivation of the upper bound~\eqref{eq_2ndmomentbound_global} is similar to the proof of~\eqref{eq_2ndmomentbound_local}. Decompose $X(t) = X^{\eta}(t) + X^{\bot}(t)$. In step~1 of~(i), we have obtained an upper bound and a lower bound for $\EE\{\|X^{\eta}(t)\|^2\}$, so it suffices to obtain an upper bound for $\EE\{\|X^{\bot}(t)\|^2\}$. Note that, from~\eqref{eq_complete_secondmoment_x_eta_lowerbound_step1} and~\eqref{eq_usedinII} 
\begin{align}\nonumber
	&\EE\{\|X^{\bot}(t+1)\|^2\} \\\nonumber
	&\le \bigg(1 - \frac32 \lambda_1\bigg) \EE\{\|X^{\eta}(t)\|^2\} + \bigg( 1 - \lambda_2 \wedge \lambda_3 - \frac12 \lambda_1\bigg) \\\nonumber
        &~\EE\{\|X^{\bot}(t)\|^2\}  + \frac{\lambda_1 }{l^{(s)}} \EE\{X(t)\}^T \tilde{M} \bfz^s + \frac{\lambda_1}{2 l^{(s)}} (z^s)^T \tilde{D}^{(s)} z^s \\ \nonumber
        &- \EE\{\| X^{\eta}(t+1)\|^2\} \\\nonumber
        &\le 
        ( 1 - \lambda_2 \wedge \lambda_3 ) \EE\{\|X^{\bot}(t)\|^2\} + \lambda_1 c_x^2 r_0 n  \Big(3 + \frac{c_l}{2}  \\ \nonumber
        &+ 4 c_s c_l + \frac{2}{r_0n} \Big)  \\ \nonumber
        &\le 
        ( 1 - \lambda_2 \wedge \lambda_3 )^t \|X^{\bot}(0)\|^2 + \lambda_1 \Big(\frac{1}{\lambda_2 \wedge \lambda_3} \wedge t\Big) c_x^2 r_0 n   \\ \label{eq_xbot_secondmoment_bound}
        &~ \Big(3 + \frac{c_l}{2}+ 4 c_s c_l+ \frac{2}{r_0n} \Big)  .
\end{align}
Therefore, the bound~\eqref{eq_2ndmomentbound_global} follows from combining the preceding bound and~\eqref{eq_complete_secondmoment_x_eta_upperbound2_step1}.
\hfill$\Box$}

\begin{lemm}\label{lem_append_extrabounds}
    Under the conditions of Lemma~\ref{lem_second_moment}, the following bounds hold,
    \begin{align*}
        | \EE\{X(t)\}^T \eta \tilde{\eta}^T z^s | &\le \tilde{l}_+^{(s)} (\|X^{\eta}(0)\| + |\zeta_1|)  \|z^s\|,\\
	| \EE\{X(t)\}^T \xi \tilde{\xi}^T \bfz^s | &\le \tilde{l}^{(s)}_- ( \|X^{\xi}(0)\| + |\zeta_2|)  \|\bfz^s\|, \\
        \Big|   \frac{\EE\{X(t)\}^T  \tilde{M} \bfz^s }{   l^{(s)}}   \Big| &\le \frac{1}{2}  (\|X(0)\|^2 + 3\|\tilde{z}^s\|^2).
    \end{align*}
\end{lemm}
\vspace{-5ex}
{\color{black}
\begin{pf}
    It follows from Lemma~\ref{lem_expression_expectation_xt} that, for all $t \in \NN$,
\begin{align}\nonumber
	&| \EE\{X(t)\}^T \eta \tilde{\eta}^T z^s | \\\nonumber
	&=
	 \Big|  \Big\{(1 - \lambda_1)^{t} \eta^T X(0) + \frac{1}{\lambda_1} [1 - (1 - \lambda_1)^{t}]  \eta^T \bar{R} z^s \Big\} \tilde{\eta}^T z^s \Big| \\\nonumber
	&\le
	(1 - \lambda_1)^{t} |\eta^T X(0) \tilde{\eta}^T z^s | +  [1 - (1 - \lambda_1)^{t}] \\\nonumber
        &\quad~ \Big| \eta^T \frac{\bar{R}}{\lambda_1} z^s  \tilde{\eta}^T z^s \Big|\\\nonumber
	&\le
	|\eta^T X(0) \tilde{\eta}^T z^s | +  \Big| \eta^T \frac{\bar{R}}{\lambda_1} z^s  \tilde{\eta}^T z^s \Big| \\\nonumber
	&\le
	\tilde{l}_+^{(s)} (\|X^{\eta}(0)\| + |\zeta_1|)  \|z^s\|,\\\nonumber
	&| \EE\{X(t)\}^T \xi \tilde{\xi}^T \bfz^s |\\\nonumber  
	&=
	  \Big|  \Big\{(1 - \lambda_2)^{t} \xi^T X(0) + \frac{1}{\lambda_2} [1 - (1 - \lambda_2)^{t}]  \xi^T \bar{R} \bfz^s \Big\} \tilde{\xi}^T \bfz^s \Big| \\ \nonumber
	&\le
	( \|X^{\xi}(0)\| + |\zeta_2|) \|\tilde{\xi}\| \|\bfz^s\|   \le
	\tilde{l}^{(s)}_- ( \|X^{\xi}(0)\| + |\zeta_2|)  \|\bfz^s\|.
\end{align}  
Again from Lemma~\ref{lem_expression_expectation_xt}, it holds that
\begin{align}\nonumber
	&\Big|   \frac{1}{   l^{(s)}} (\EE\{X(t)\})^T  \tilde{M} \bfz^s   \Big| \\\nonumber
	&=
	\Big| (1 - \lambda_1)^{t}  \eta^T X(0) \eta^T \frac{\tilde{M}}{l^{(s)}} \bfz^s +  [1 - (1 - \lambda_1)^{t}] \eta^T \frac{\bar{R}}{\lambda_1} \bfz^s\\\nonumber
        &~  \eta^T \frac{\tilde{M}}{l^{(s)}} \bfz^s + (1 - \lambda_2)^{t}  \xi^T X(0) \xi^T \frac{\tilde{M}}{l^{(s)}} \bfz^s + \frac{\lambda_1}{\lambda_2} [1- (1  \\\nonumber
	& - \lambda_2)^{t}]  \xi^T \frac{\bar{R}}{\lambda_1} \bfz^s \xi^T \frac{\tilde{M}}{l^{(s)}} \bfz^s + (1 - \lambda_3)^{t} \sum_{i = 3}^{r_0n}  (w^{(i)})^TX(0)\\\nonumber
        & (w^{(i)})^T \frac{\tilde{M}}{l^{(s)}} \bfz^s + \frac{\lambda_1}{\lambda_3} [1 - (1 - \lambda_3)^{t}] \sum_{i = 3}^{r_0n}  (w^{(i)})^T \frac{\bar{R}}{\lambda_1} \bfz^s\\\nonumber
	&  (w^{(i)})^T \frac{\tilde{M}}{l^{(s)}} \bfz^s \Big| \\\nonumber
	&\le
	\Big|  \eta^T X(0) \eta^T \frac{\tilde{M}}{l^{(s)}} \bfz^s \Big| +  \Big|  \eta^T \frac{\bar{R}}{\lambda_1} \bfz^s \eta^T \frac{\tilde{M}}{l^{(s)}} \bfz^s \Big| + \Big|  \xi^T X(0) \xi^T  \\\nonumber
        &~\frac{\tilde{M}}{l^{(s)}} \bfz^s \Big| + \Big|   \xi^T \frac{\bar{R}}{\lambda_1} \bfz^s \xi^T \frac{\tilde{M}}{l^{(s)}} \bfz^s \Big| + \Big| \sum_{i = 3}^{r_0n}  (w^{(i)})^TX(0) (w^{(i)})^T \\\nonumber
        &~\frac{\tilde{M}}{l^{(s)}} \bfz^s \Big| + \Big| \sum_{i = 3}^{r_0n}  (w^{(i)})^T \frac{\bar{R}}{\lambda_1} \bfz^s (w^{(i)})^T \frac{\tilde{M}}{l^{(s)}} \bfz^s \Big| \\\nonumber
	&\le 
	( \|X^{\eta}(0)\|~|\zeta_1| + |\zeta_1|^2 + \|X^{\xi}(0)\|~|\zeta_2| + |\zeta_2|^2 \\\label{eq_complete_secondmoment_x_gamma_step3_temp_M1}
        &+ \|X^{\Gamma}(0)\|~|\zeta_3| + |\zeta_3|^2 ) \\\nonumber
	&\le
	\frac{1}{2} [ \|X^{\eta}(0)\|^2   +  \|X^{\xi}(0)\|^2 + \|X^{\Gamma}(0)\|^2 + 3(|\zeta_1|^2 \\\nonumber
        &  + |\zeta_2|^2  +   |\zeta_3|^2) ] \tag{from $2ab \le a^2 + b^2$, $\forall a,b\in \RR$}\\\nonumber
	&=
	\frac{1}{2}  (\|X(0)\|^2 + 3\|\tilde{z}^s\|^2). 
\end{align}
Here, the last equation follows from Lemma~\ref{lem_eigen}~(ii), and~\eqref{eq_complete_secondmoment_x_gamma_step3_temp_M1} is obtained from the following fact,
\begin{align*}
	&\Big| \sum_{i = 3}^{r_0n}  (w^{(i)})^TX(0) (w^{(i)})^T \frac{\tilde{M}}{l^{(s)}} \bfz^s \Big| \\
        &\le \Big(\sum_{i = 3}^{r_0n}  ((w^{(i)})^TX(0))^2 \Big)^{\frac12} \Big( \sum_{i = 3}^{r_0n}  \Big((w^{(i)})^T \frac{\tilde{M}}{l^{(s)}} \bfz^s \Big)^2 \Big)^{\frac12} \\
	& = \Big\| \sum_{i = 3}^{r_0n} w^{(i)} (w^{(i)})^T X(0) \Big\|~  \Big\| \sum_{i = 3}^{r_0n} w^{(i)} (w^{(i)})^T \frac{\tilde{M}}{l^{(s)}} \bfz^s \Big\| \\
	&= \|X^{\Gamma}(0) \|~|\zeta_3|,
\end{align*}
where the first inequality follows from the Cauchy-Schwarz inequality, and the first equation is from the orthogonality of $w^{(i)}$, $3\le i \le r_0n$.  \hfill$\Box$ 
\end{pf}
}

{\color{black}\section{Proof of Theorem~\ref{thm_deviationInterval}}\label{sec_proofs_thmdevInt}
Similar to Theorem~\ref{thm_deviationFromEt}, to prove~(i) it suffices to bound the probability $\PP\{|\mtcs(X(t),X^\eta(0)+X^\xi(0),\eps)| \ge \delta r_0n \}$. From Lemma~\ref{lem_expression_expectation_xt} and the Bernoulli inequality we have that
\begin{align}\nonumber
	& \EE\{ \|X(t) - (X^{\eta}(0)+X^{\xi}(0)) \|^2\} \\\nonumber
	&= 
	\EE\{\|X(t)\|^2\} + \|X^{\eta}(0)+X^{\xi}(0) \|^2- 2 \EE\{X(t)\}^T \\\nonumber
        &~(X^{\eta}(0)+X^{\xi}(0)) \\\nonumber
	&=
	\EE\{\|X(t)\|^2\} +  \|X^{\eta}(0)\|^2 + \|X^{\xi}(0) \|^2 - 2 (1-\lambda_1)^t \\ \nonumber
        &~\|X^{\eta}(0)\|^2 - [1 - (1-\lambda_1)^t] \zeta_1 \eta^T X(0)  - 2 (1-\lambda_2)^t \\\nonumber
	&~ \|X^{\xi}(0)\|^2  - \frac{\lambda_1}{\lambda_2} [1-(1-\lambda_2)^t]  \zeta_2 \xi^T X(0) \\\nonumber
	&\le
	\EE\{\|X(t)\|^2\} + [1 -2 (1-\lambda_1)^t ]\|X^{\eta}(0)\|^2 \\ \nonumber
        &+ [1 - 2 (1-\lambda_2)^t ] \|X^{\xi}(0) \|^2  +  2\lambda_1 t |\zeta_1|\|X^{\eta}(0)\| \\ \nonumber
        &+  2\lambda_1 t |\zeta_2|\|X^{\xi}(0)\| 
        \\\nonumber
        &\le 
        \EE\{\|X(t)\|^2\} + [1 -2 (1-\lambda_1)^t ]\|X^{\eta}(0)\|^2 \\ \nonumber
        &+ [1 - 2 (1-\lambda_2)^t ] \|X^{\xi}(0) \|^2  +  2\lambda_1 t c_x^2r_0n \\ \nonumber
	&\le
        (1-\lambda_3)^t \|X^\Gamma(0)\|^2 + \lambda_1 t c_x^2r_0n \Big( 6 + \frac{c_l}{2} + 16 c_s c_l \\ \nonumber
        &+\frac{3c_l + 23}{2r_0n} \Big) + \Big( \frac{\lambda_2}{\lambda_3} + \lambda_2 t\Big) (\|X^\xi(0)\|^2 + c_x^2),
\end{align}
where the last inequality follows from~\eqref{eq_2ndmomentbound_local} with $1/\lambda_3 \le t \le 1/\lambda_2$. Hence from~\eqref{eq_Sbound}, 
\begin{align*}
    &\PP\{|\mtcs(X(t),X^\eta(0)+X^\xi(0),\eps)| \ge \delta r_0n \} \\
    &\le \frac{1}{\eps^2\delta} \Big[ (1-\lambda_3)^t + (3+C_{11})\lambda_1 t + C_{12} \Big( \frac{\lambda_2}{\lambda_3} + \lambda_2 t\Big) \Big],
\end{align*}
and the conclusion of~(i) follows.
The second part of the theorem can be derived from similar calculations.
}

\section{Proof of Theorem~\ref{thm_prob3}}\label{sec_proofs_thm_prob3}
In this section, we denote the maximum-column-sum, spectral, and maximum-row-sum norm by $\|\cdot\|_1$, $\|\cdot\|$, and $\|\cdot\|_{\infty}$, respectively.
Let $\{[Q^{\prime}(t)~ U^{\prime}(t)], t\in \mathbb{N}\}$ be an i.i.d. sequence having the same distribution as and independent of the original sequence $\{[Q(t)~ U(t)], t\in \mathbb{N}\}$, where $U(t) \Let R(t)z^s$. Denote $\Phi_{Q^{\prime}}(s,t) = Q^{\prime}(t)\cdots Q^{\prime}(s)$, $\overleftarrow{\Phi}_{Q^{\prime}}(s,t) = Q^{\prime}(s)\cdots Q^{\prime}(t)$, and $\Phi_{Q^{\prime}}(t+1,t) = \overleftarrow{\Phi}_{Q^{\prime}}(t+1,t) = I$, for all $t\ge s\ge 0$. Define 
\begin{align*}
    \tilde{X}(t) &:= \Phi_{Q^{\prime}}(0,t) X(0)  + \sum_{i=0}^t \Phi_{Q^{\prime}}(t+1-i,t) U^{\prime}(t-i),\\
    \tilde{X}^*(t) &:= \sum_{i=0}^t \Phi_{Q^{\prime}}(t+1-i,t) U^{\prime}(t-i) + \Phi_{Q^{\prime}}(0,t) \\
    &~\sum_{i=t+1}^{\infty} \overleftarrow{\Phi}_{Q^{\prime}}(t+1,i-1) U^{\prime}(i),
\end{align*}
so $\tilde{X}(t)$ and $\tilde{X}^{*}(t)$ have the same distribution as $X(t)$ and $\pi$ respectively (the latter correspondence is because $\pi$ has the same distribution as $\sum_{i=0}^{\infty} \overleftarrow{\Phi}_{Q}(0,i-1) U(i)$ \cite{xing2023community,acemouglu2013opinion}). Hence, the following observation yields the conclusion.
\begin{align*}
    &~d_W(X(t),\pi)
    \le \mathbb{E}\{\|\tilde{X}(t)-\tilde{X}^{*}(t)\|\} \\
    &
    = \mathbb{E}\Big\{ \Big\| \Phi_{Q^{\prime}}(0,t) X(0)   - \Phi_{Q^{\prime}}(0,t)  \\
    &\quad~ \sum\nolimits_{i=t+1}^{\infty} \overleftarrow{\Phi}_{Q^{\prime}}(t+1,i-1)U^{\prime}(i) \Big\|\Big\}\\
    &
    \le c_x\sqrt{r_0n} \mathbb{E}\{\| \Phi_{Q^{\prime}}(0,t) \|\}  \\
    &\quad~ + \frac12 c_x \sum\nolimits_{i=t+1}^{\infty} \mathbb{E}\{\|\Phi_{Q^{\prime}}(0,t)\overleftarrow{\Phi}_{Q^{\prime}}(t+1,i-1)\|\}\\
    &
    \le \frac12 c_x r_0 n \Big(2 \mathbb{E}\{\| \Phi_{Q^{\prime}}(0,t) \|_1\} \\
    &\quad~ + \sum\nolimits_{i=t+1}^{\infty} \mathbb{E}\{\|\Phi_{Q^{\prime}}(0,t) \overleftarrow{\Phi}_{Q^{\prime}}(t+1,i-1)\|_1\}\Big)\\
    &\le 
    c_x (r_0 n)^{\frac52} \Big(1 + \frac{1}{2\lambda_1}\Big) (1-\lambda_1)^{t+1},
\end{align*}
where the first inequality follows from the definition of the Wasserstein metric, the second inequality from Assumption~\ref{asmp_2}, and the last inequality is obtained by noticing the following fact
\begin{align*}
	&\mathbb{E}\{ \| \overleftarrow{\Phi}_Q(0,t) \|_{1} \}  
	\le 
	\sum_{1 \le j \le r_0n} \sum_{1 \le i \le r_0n} \mathbb{E} \{ |[ \overleftarrow{\Phi}_Q(0,t) ]_{ij}| \}\\
        &\le
	r_0 n \| \mathbb{E} \{ \overleftarrow{\Phi}_Q(0,t) \} \|_{\infty} = r_0n \|\bar{Q}^{t+1}\|_{\infty} \\
        &\le 
	(r_0 n)^{\frac32} (1-\lambda_1)^{t+1}.
\end{align*}

\bibliographystyle{ieeetr}
\bibliography{bibliography.bib}

\begin{thebibliography}{10}

\bibitem{proskurnikov2018tutorial}
A.~V. Proskurnikov and R.~Tempo, ``A tutorial on modeling and analysis of
  dynamic social networks. {P}art {II},'' {\em Annual Reviews in Control},
  vol.~45, pp.~166--190, 2018.

\bibitem{fortunato2016community}
S.~Fortunato and D.~Hric, ``Community detection in networks: A user guide,''
  {\em Physics Reports}, vol.~659, pp.~1--44, 2016.

\bibitem{conover2011political}
M.~Conover, J.~Ratkiewicz, M.~Francisco, B.~Gon{\c{c}}alves, F.~Menczer, and
  A.~Flammini, ``Political polarization on twitter,'' in {\em Proceedings of
  the International AAAI Conference on Web and Social Media}, vol.~5,
  pp.~89--96, 2011.

\bibitem{cota2019quantifying}
W.~Cota, S.~C. Ferreira, R.~Pastor-Satorras, and M.~Starnini, ``Quantifying
  echo chamber effects in information spreading over political communication
  networks,'' {\em EPJ Data Science}, vol.~8, no.~1, pp.~1--13, 2019.

\bibitem{schaub2020blind}
M.~T. Schaub, S.~Segarra, and J.~N. Tsitsiklis, ``Blind identification of
  stochastic block models from dynamical observations,'' {\em SIAM Journal on
  Mathematics of Data Science}, vol.~2, no.~2, pp.~335--367, 2020.

\bibitem{xing2023community}
Y.~Xing, X.~He, H.~Fang, and K.~H. Johansson, ``Community structure recovery
  and interaction probability estimation for gossip opinion dynamics,'' {\em
  Automatica}, vol.~154, p.~111105, 2023.

\bibitem{cheng2018model}
X.~Cheng, Y.~Kawano, and J.~M. Scherpen, ``Model reduction of multiagent
  systems using dissimilarity-based clustering,'' {\em IEEE Transactions on
  Automatic Control}, vol.~64, no.~4, pp.~1663--1670, 2018.

\bibitem{castellano2009statistical}
C.~Castellano, S.~Fortunato, and V.~Loreto, ``Statistical physics of social
  dynamics,'' {\em Reviews of Modern Physics}, vol.~81, no.~2, p.~591, 2009.

\bibitem{flache2017models}
A.~Flache, M.~M{\"a}s, T.~Feliciani, E.~Chattoe-Brown, G.~Deffuant, S.~Huet,
  and J.~Lorenz, ``Models of social influence: Towards the next frontiers,''
  {\em Journal of Artificial Societies and Social Simulation}, vol.~20, no.~4,
  2017.

\bibitem{de2019learning}
A.~De, S.~Bhattacharya, P.~Bhattacharya, N.~Ganguly, and S.~Chakrabarti,
  ``Learning linear influence models in social networks from transient opinion
  dynamics,'' {\em ACM Transactions on the Web}, vol.~13, no.~3, pp.~1--33,
  2019.

\bibitem{friedkin2021group}
N.~E. Friedkin, A.~V. Proskurnikov, and F.~Bullo, ``Group dynamics on
  multidimensional object threat appraisals,'' {\em Social Networks}, vol.~65,
  pp.~157--167, 2021.

\bibitem{kozitsin2023opinion}
I.~V. Kozitsin, ``Opinion dynamics of online social network users: {A}
  micro-level analysis,'' {\em The Journal of Mathematical Sociology}, vol.~47,
  no.~1, pp.~1--41, 2023.

\bibitem{degroot1974reaching}
M.~H. DeGroot, ``Reaching a consensus,'' {\em Journal of the American
  Statistical Association}, vol.~69, no.~345, pp.~118--121, 1974.

\bibitem{friedkin1990social}
N.~E. Friedkin and E.~C. Johnsen, ``Social influence and opinions,'' {\em
  Journal of Mathematical Sociology}, vol.~15, no.~3-4, pp.~193--206, 1990.

\bibitem{hegselmann2002opinion}
R.~Hegselmann and U.~Krause, ``Opinion dynamics and bounded confidence models,
  analysis, and simulation,'' {\em Journal of Artificial Societies and Social
  Simulation}, vol.~5, no.~3, 2002.

\bibitem{deffuant2000mixing}
G.~Deffuant, D.~Neau, F.~Amblard, and G.~Weisbuch, ``Mixing beliefs among
  interacting agents,'' {\em Advances in Complex Systems}, vol.~3, no.~01n04,
  pp.~87--98, 2000.

\bibitem{shi2019dynamics}
G.~Shi, C.~Altafini, and J.~S. Baras, ``Dynamics over signed networks,'' {\em
  SIAM Review}, vol.~61, no.~2, pp.~229--257, 2019.

\bibitem{abelson1964mathematical}
R.~P. Abelson, ``Mathematical models of the distribution of attitudes under
  controversy,'' {\em Contributions to Mathematical {P}sychology}, 1964.

\bibitem{banisch2012agent}
S.~Banisch, R.~Lima, and T.~Ara{\'u}jo, ``Agent based models and opinion
  dynamics as {M}arkov chains,'' {\em Social Networks}, vol.~34, no.~4,
  pp.~549--561, 2012.

\bibitem{chowell2016mathematical}
G.~Chowell, L.~Sattenspiel, S.~Bansal, and C.~Viboud, ``Mathematical models to
  characterize early epidemic growth: A review,'' {\em Physics of Life
  Reviews}, vol.~18, pp.~66--97, 2016.

\bibitem{noorazar2020classical}
H.~Noorazar, K.~R. Vixie, A.~Talebanpour, and Y.~Hu, ``From classical to modern
  opinion dynamics,'' {\em International Journal of Modern Physics C}, vol.~31,
  no.~07, p.~2050101, 2020.

\bibitem{banisch2010empirical}
S.~Banisch and T.~Ara{\'u}jo, ``On the empirical relevance of the transient in
  opinion models,'' {\em Physics Letters A}, vol.~374, no.~31-32,
  pp.~3197--3200, 2010.

\bibitem{hill2013quickly}
S.~J. Hill, J.~Lo, L.~Vavreck, and J.~Zaller, ``How quickly we forget: The
  duration of persuasion effects from mass communication,'' {\em Political
  Communication}, vol.~30, no.~4, pp.~521--547, 2013.

\bibitem{lorenz2006consensus}
J.~Lorenz, ``Consensus strikes back in the {H}egselmann-{K}rause model of
  continuous opinion dynamics under bounded confidence,'' {\em Journal of
  Artificial Societies and Social Simulation}, vol.~9, no.~1, 2006.

\bibitem{barbillon2015network}
P.~Barbillon, M.~Thomas, I.~Goldringer, F.~Hospital, and S.~Robin, ``Network
  impact on persistence in a finite population dynamic diffusion model:
  {A}pplication to an emergent seed exchange network,'' {\em Journal of
  Theoretical Biology}, vol.~365, pp.~365--376, 2015.

\bibitem{dietrich2016transient}
F.~Dietrich, S.~Martin, and M.~Jungers, ``Transient cluster formation in
  generalized {H}egselmann-{K}rause opinion dynamics,'' in {\em European
  Control Conference}, pp.~531--536, 2016.

\bibitem{xiong2017modeling}
F.~Xiong, Y.~Liu, and J.~Cheng, ``Modeling and predicting opinion formation
  with trust propagation in online social networks,'' {\em Communications in
  Nonlinear Science and Numerical Simulation}, vol.~44, pp.~513--524, 2017.

\bibitem{s2022finite}
S.~S. Shree, C.~Avhishek, and J.~Krishna, ``Finite time bounds for stochastic
  bounded confidence dynamics,'' {\em arXiv preprint arXiv:2212.13387}, 2022.

\bibitem{festinger1949analysis}
L.~Festinger, ``The analysis of sociograms using matrix algebra,'' {\em Human
  Relations}, vol.~2, no.~2, pp.~153--158, 1949.

\bibitem{newman2004finding}
M.~E. Newman and M.~Girvan, ``Finding and evaluating community structure in
  networks,'' {\em Physical Review E}, vol.~69, no.~2, p.~026113, 2004.

\bibitem{abbe2017community}
E.~Abbe, ``Community detection and stochastic block models: {R}ecent
  developments,'' {\em The Journal of Machine Learning Research}, vol.~18,
  no.~1, pp.~6446--6531, 2017.

\bibitem{gargiulo2010opinion}
F.~Gargiulo and S.~Huet, ``Opinion dynamics in a group-based society,'' {\em
  Europhysics Letters}, vol.~91, no.~5, p.~58004, 2010.

\bibitem{fennell2021generalized}
S.~C. Fennell, K.~Burke, M.~Quayle, and J.~P. Gleeson, ``Generalized mean-field
  approximation for the {D}effuant opinion dynamics model on networks,'' {\em
  Physical Review E}, vol.~103, no.~1, p.~012314, 2021.

\bibitem{baumann2020laplacian}
F.~Baumann, I.~M. Sokolov, and M.~Tyloo, ``A {L}aplacian approach to stubborn
  agents and their role in opinion formation on influence networks,'' {\em
  Physica A: Statistical Mechanics and its Applications}, vol.~557, p.~124869,
  2020.

\bibitem{si2009opinion}
X.~Si, Y.~Liu, and Z.~Zhang, ``Opinion dynamics in populations with implicit
  community structure,'' {\em International Journal of Modern Physics C},
  vol.~20, no.~12, pp.~2013--2026, 2009.

\bibitem{como2016local}
G.~Como and F.~Fagnani, ``From local averaging to emergent global behaviors:
  The fundamental role of network interconnections,'' {\em Systems \& Control
  Letters}, vol.~95, pp.~70--76, 2016.

\bibitem{boyd2006randomized}
S.~Boyd, A.~Ghosh, B.~Prabhakar, and D.~Shah, ``Randomized gossip algorithms,''
  {\em IEEE Transactions on Information Theory}, vol.~52, no.~6,
  pp.~2508--2530, 2006.

\bibitem{fagnani2008randomized}
F.~Fagnani and S.~Zampieri, ``Randomized consensus algorithms over large scale
  networks,'' {\em IEEE Journal on Selected Areas in Communications}, vol.~26,
  no.~4, pp.~634--649, 2008.

\bibitem{acemouglu2013opinion}
D.~Acemo{\u{g}}lu, G.~Como, F.~Fagnani, and A.~Ozdaglar, ``Opinion fluctuations
  and disagreement in social networks,'' {\em Mathematics of Operations
  Research}, vol.~38, no.~1, pp.~1--27, 2013.

\bibitem{holme2006nonequilibrium}
P.~Holme and M.~E. Newman, ``Nonequilibrium phase transition in the coevolution
  of networks and opinions,'' {\em Physical Review E}, vol.~74, no.~5,
  p.~056108, 2006.

\bibitem{biswas2012disorder}
S.~Biswas, A.~Chatterjee, and P.~Sen, ``Disorder induced phase transition in
  kinetic models of opinion dynamics,'' {\em Physica A: Statistical Mechanics
  and its Applications}, vol.~391, no.~11, pp.~3257--3265, 2012.

\bibitem{shi2016evolution}
G.~Shi, A.~Proutiere, M.~Johansson, J.~S. Baras, and K.~H. Johansson, ``The
  evolution of beliefs over signed social networks,'' {\em Operations
  Research}, vol.~64, no.~3, pp.~585--604, 2016.

\bibitem{xing2022what}
Y.~Xing and K.~H. Johansson, ``What is the expected transient behavior of
  opinion evolution for two communities?,'' {\em arXiv preprint
  arXiv:2304.12495}, 2023.

\bibitem{chung2011spectra}
F.~Chung and M.~Radcliffe, ``On the spectra of general random graphs,'' {\em
  The Electronic Journal of Combinatorics}, p.~P215, 2011.

\end{thebibliography}





%
%
%

\end{document}